\documentclass{aa}
\usepackage{graphicx}
\usepackage{textcomp}
\usepackage{lscape}
\usepackage{subfigure}
\usepackage{natbib}
\usepackage[varg]{txfonts}
\usepackage{longtable}
\usepackage{booktabs}
\usepackage[normalem]{ulem} 
\usepackage{epstopdf}
\usepackage{xcolor}
\usepackage{soul}
\usepackage{makecell}
\usepackage{multicol,tabularx,capt-of}
\usepackage{scalerel}
\usepackage{tikz}
\usepackage{lineno}

\usetikzlibrary{svg.path}

\defcitealias{Sanchez-Saez2023}{SS23}

\definecolor{orcidlogocol}{HTML}{A6CE39}
\tikzset{
  orcidlogo/.pic={
    \fill[orcidlogocol] svg{M256,128c0,70.7-57.3,128-128,128C57.3,256,0,198.7,0,128C0,57.3,57.3,0,128,0C198.7,0,256,57.3,256,128z};
    \fill[white] svg{M86.3,186.2H70.9V79.1h15.4v48.4V186.2z}
                 svg{M108.9,79.1h41.6c39.6,0,57,28.3,57,53.6c0,27.5-21.5,53.6-56.8,53.6h-41.8V79.1z M124.3,172.4h24.5c34.9,0,42.9-26.5,42.9-39.7c0-21.5-13.7-39.7-43.7-39.7h-23.7V172.4z}
                 svg{M88.7,56.8c0,5.5-4.5,10.1-10.1,10.1c-5.6,0-10.1-4.6-10.1-10.1c0-5.6,4.5-10.1,10.1-10.1C84.2,46.7,88.7,51.3,88.7,56.8z};
  }
}

\newcommand\orcidicon[1]{\href{https://orcid.org/#1}{\mbox{\scalerel*{
\begin{tikzpicture}[yscale=-1,transform shape]
\pic{orcidlogo};
\end{tikzpicture}
}{|}}}}

\usepackage{hyperref} 
\hypersetup{
    colorlinks=true,
    citecolor=blue,
    linkcolor=blue,
    urlcolor=blue,
    }
\makeatletter
\renewcommand*\aa@pageof{, page \thepage{} of \pageref*{LastPage}}
\makeatother

\bibpunct{(}{)}{;}{a}{}{,}

\definecolor{cadmiumred}{rgb}{0.89, 0.0, 0.13}

\definecolor{ste}{rgb}{0., 0.26, 0.15}

\begin{document} 

   \title{Seeing through the dust: Unraveling near-infrared variability in type 2 active galactic nuclei}
      \titlerunning{NIR variability of type 2 AGNs}
   \authorrunning{ Cerón-Meneses et al.}

   \author{J. Cerón-Meneses\inst{1,2,3}, P. Arévalo \inst{1,2,3\orcidicon{0000-0001-5675-6323}}, P. Sánchez-Sáez \inst{4}, W. Yu \inst{5 \orcidicon{0000-0003-1262-2897}}, P. Lira \inst{6,3}, B. Milvang-Jensen\inst{7,8}}

\institute{Instituto de Física y Astronomía, Universidad de Valparaíso, Gran Bretaña 1111, Valparaíso, Chile\\
             \email{jhon.ceron@postgrado.uv.cl}
              \and
Millennium Nucleus on Transversal Research and Technology to Explore Supermassive Black Holes (TITANS)
\and Millennium Institute of Astrophysics (MAS), Nuncio Monseñor Sótero Sanz 100, Providencia, Santiago, Chile
\and  
European Southern Observatory, Karl-Schwarzschild-Strasse 2, 85748 Garching bei München, Germany
\and 
Department of Physics \& Astronomy, Bishop's University, 2600 rue College, Sherbrooke, QC, J1M 1Z7, Canada
\and 
Departamento de Astronomía, Universidad de Chile, Casilla 36D, Santiago, Chile
\and 
Cosmic Dawn Center (DAWN)
\and 
Niels Bohr Institute, University of Copenhagen, Jagtvej 128, 2200 Copenhagen N, Denmark
}

   \date{}
  \abstract
   {Near-infrared (NIR) variability studies of active galactic nuclei (AGNs) are still limited, as long-term multiepoch monitoring in the NIR is observationally challenging. The depth, wavelength coverage, and 14-year temporal baseline of UltraVISTA make it one of the few surveys capable of providing a detailed characterization of AGN variability in this regime.}
   {We aim to quantify the NIR variability of known AGNs in the COSMOS field and to investigate the physical origin of variability in type 2 AGNs. In particular, we examine how NIR variability can help clarify the discrepancies between optical and X-ray classifications.
}
   {Using the 14-year multiepoch UltraVISTA DR6 dataset in the $YJHK_s$ bands, we constructed calibrated NIR light curves and quantified their variability through a set of metrics. Active galactic nucleus-like stochastic variability was identified by modeling the light curves with a damped random walk (DRW) process.
}
   {We find that $\sim$7-17\% of the 533 type 2 AGNs are variable in the NIR, with variability fractions increasing toward $K_s$, where the dusty torus dominates the emission. Based on the wavelength dependence of the DRW variability amplitude, we classify variable type 2 AGNs into disk-dominated, torus-dominated, and highly obscured groups. About one third of the X-ray unobscured (XR I) type 2 AGNs are variable in the NIR, consistent with misclassified weak type 1 or ``true type 2'' AGNs. On the other hand, 21.4\% (30/140) of the X-ray obscured (XR II) type 2 AGNs show detectable variability in the NIR, most of them only in $H$ or $K_s$, consistent with obscuration of the bluer (accretion disk) bands. Type 2 AGNs without X-ray counterparts (165) show the smallest fraction ($3.6$\%) of variable objects.
}
   {NIR variability provides an effective and independent diagnostic for confirming optical classifications and for identifying weak or misclassified type 1 AGNs in deep extragalactic surveys.
}

   \maketitle

\section{Introduction}\label{section:intro}

Our current understanding of accreting supermassive black holes at the centers of galaxies includes an optically thick accretion disk that shines in the optical/UV as the primary source of luminosity \citep{Salpeter1964, Lynden-Bell1969}. This emission is variable, a feature that has been widely used to map the unresolvable inner regions of many active galactic nuclei (AGNs; \citealt[and references therein)]{Cackett2021}. The distance from the accretion disk to the gas structure responsible for the emission of broad lines has been measured through reverberation mapping, i.e., through the time delay between fluctuations of the primary continuum emission and the reprocessed line emission \citep[e.g.,][]{Blandford1982}. In a similar way, the distance to the dusty torus has also been measured from the delays between the optical and near-infrared (NIR) variations \citep[e.g.,][]{Minezaki2004}. Given that the hottest dust, closest to the central engine, can only be heated up to its sublimation temperature, the inner edge of the dusty torus emits thermal radiation that peaks at NIR wavelengths. While the $J$, $H$, and $K_s$ bands all probe the $\sim$1–2 $\mu$m regime, the contribution from the accretion disk decreases rapidly toward longer wavelengths. As a result, the $K_s$ band provides a more direct probe of the hot dust emission of the torus, and minimizes contamination from the optical/UV disk emission \citep{Barvainis1992}.

These reverberation mapping campaigns prove the direct connection between optical continuum variations and reprocessed broad line and NIR variability, and that the broad line region (BLR) effectively resides within the dusty torus \citep{Sugauma2006}, as predicted by the unified model of AGNs \citep{Antonucci1985,Urry1995}. In this scenario, type 1 AGNs, i.e., those with broad emission lines and typically a blue continuum in their optical/UV spectra, present an unobscured view of the disk and BLR. In contrast, type 2 AGNs, where the broad lines and, typically, the blue continuum are not visible, would have the disk and BLR obscured by a dusty torus blocking our line of sight. A corollary of this model is that the optical variability should be strongly suppressed in type 2 AGNs, given that the variable disk emission is obscured. The NIR variability, however, could still appear in type 2s, if the inner torus was visible either directly or through the dust, which is less opaque at NIR wavelengths than in the optical range. Therefore, a clear sign of obscuration is significant variability in the NIR with no contemporaneous optical variations.

A detailed search for optical variations in type 2 AGNs was conducted by \citet[see also \citealt{Barth2014A}]{LopezNavas2022, LopezNavas2023}. They used 2.5-year light curves from the \textit{Zwicky} Transient Facility for over 15,000 type 2 AGNs. Their findings include negligible variability in the type 2 sample in general, but significant variability was detected in a subsample classified as weak type 1 by \citet{Oh2015} and \citet{Liu2019}; this variability was detected via detailed spectral modeling that detected weak but significant broad emission lines. The emerging picture of optical variability therefore points to no variability in type 2 AGNs, weak variability in weakly emitting but unobscured AGNs, and stronger variability in strongly emitting, traditional type 1 AGNs.  

Type 1 AGNs are typically more variable in bluer bands \citep[e.g.,][]{MacLeod2010, Simm2016, Sanchez2017, Li2018, Sanchez2018, Yu2022, Patel2025}. When the observed band transitions into the wavelengths dominated by the torus, however, the variability amplitude can once again increase toward longer wavelengths \citep[e.g.,][]{Lira2015}. This can be understood considering that the dust emission responds to variations in the whole disk, which can be dominated by large amplitude fluctuations in the UV, far from the observed, less variable optical bands. According to this ratio of disk-to-dust variability amplitude, \citet{Lira2011, Lira2024} suggest a differentiation between disk-dominated variations, where longer wavelengths show consistently less variability, and torus-dominated variations, where the amplitude of variability drops toward longer wavelengths up to the $J$ or $H$ band and then increases again. 

An additional diagnostics on the obscuration toward the line of sight to the AGNs is provided by observations in the X-ray regime. The shape of the X-ray spectrum enables the column density of the obscuring gas to be measured. For large column densities, this gas is likely located in the obscuring torus, whose dust should also obscure the optical light from the disk and BLR. The relation between X-ray obscuration and optical type 1/type 2 classification is good but not perfect \citep{Merloni2014, Koss2017, Boorman2025}, in particular for the sample in the COSMOS field, which is the subject of this paper. As discussed by these authors, mismatched objects such as X-ray unobscured type 2 AGNs could correspond to incorrect optical classifications due to host galaxy dilution. As such, these unobscured type 2 AGNs could show variability in the optical and NIR bands.

Active galactic nucleus-like variability can be modeled as a stochastic process commonly described by a damped random walk (DRW) model  \citep[e.g.,][]{Kelly2009, MacLeod2010, Kolzlowski2010}. The DRW process corresponds to the lowest-order continuous-time autoregressive moving-average model \citep{Brockwell2001, Kelly2014}, CARMA(1,0), and has been widely used to describe UV/optical variability in AGNs. In this framework, the power spectral density (PSD) follows a broken power-law shape ($1/f^{\alpha}$), where the exponent $\alpha$ is approximately zero at low frequencies and approaches two at higher frequencies. The frequency at which this change occurs is known as the break frequency, which corresponds to a characteristic timescale ($\tau_{\rm DRW}$). Several studies have successfully applied the DRW model to characterize AGN light curves and have reported correlations between the variability properties of this model and the fundamental parameters of AGNs such as black hole mass, accretion rate, and luminosity \citep[e.g.,][]{MacLeod2010, Simm2016, Sanchez2018, Burke2021,Stone2022}. In addition, DRW parameters are commonly used as features to classify variable sources in light-curve classifiers such as the Automatic Learning for the Rapid Classification of Events (ALeRCE) broker \citep{Sanchez2021}. Despite the broad use of the DRW model, deviations from this behavior have also been reported, particularly on timescales much shorter than the characteristic damping timescale, where the power spectral slope appears to be steeper than in the DRW model \citep[e.g.,][]{Mushotzky2011,Kasliwal2015,Smith2018, Petrecca2024, Arevalo2024,Yu2025a}.

Our study focuses on the detection and characterization of variability signatures associated with type 2 AGNs within the COSMOS field based on the NIR data. The primary objective is to quantify the frequency of these variable events and classify the detected sources into categories dominated by the accretion disk, the dusty torus, or those sources that are highly obscured. Furthermore, we establish a direct connection between these variability classifications and their corresponding X-ray obscuration. To achieve this, we conducted an exhaustive re-analysis of the COSMOS field observations obtained by the UltraVISTA survey \citep{McCracken2012} in the $Y$, $J$, $H$, and $K_s$ bands, and integrated all available data. This work builds upon the methodology presented in \citet{Sanchez2017} for the construction of NIR light curves. Crucially, the parameters of the sample are significantly expanded and incorporate twice the observed area, longer light curves, and improved statistics used in the search for AGN-like variability.

This paper is organized as follows. In Sect. \ref{section:data} we describe the UltraVISTA DR6 multiepoch dataset used in this work. Section \ref{section:light curve construction} explains the photometric calibration and the construction and cleaning of the $YJHK_s$ light curves. In Sect. \ref{section:agn sample} we show how the AGN sample was assembled. Section \ref{section:features} presents the variability features used in the analysis and the selection criteria for stochastic AGN-like variability. Section \ref{section:results} presents the main results, including the variability fractions of the different AGN classes and the characterization of their NIR variability amplitudes. In Sect. \ref{section:discussion} we discuss the physical implications of our findings, a comparison with previous work, and the connection between optical and X-ray classifications. Finally, Sect. \ref{section:conclusions} summarizes our conclusions. The photometry reported is in the AB system.

\section{Data}\label{section:data}

We used data from the sixth and final UltraVISTA data release (DR6\footnote{See the release description document: \url{https://www.eso.org/rm/api/v1/public/releaseDescriptions/221}}; \citeauthor{McCracken2012} \citeyear{McCracken2012}), an ultra-deep NIR survey in the $YJHK_s$ and NB118 bands, covering the central region of the COSMOS field. The observations, carried out with the VIRCAM instrument on the VISTA telescope at Paranal \citep{Emerson2006, Dalton2006}, span 14 seasons from December 2009 to March 2023, providing multiepoch coverage within each season. The VIRCAM footprint on the sky consists of 16 widely spaced detectors. By stepping the array along right ascension and declination, a contiguous image is obtained, defining four ``deep'' and four ``ultra-deep'' stripes. To ensure uniform sky coverage in our analysis, we excluded the NB118 narrow-band filter, which is only available across the ultra-deep stripes. Although DR6 also includes NB118 data in the deep stripes, these correspond to a single observing season and therefore do not provide homogeneous temporal coverage. The pixel scale of the images is $0.34''\,\mathrm{pixel^{-1}}$.

To construct our light curves we used individual observation block (OB) stacks, corresponding to 0.5 or 1 hr of exposure, produced by the Cambridge Astronomy Survey Unit (CASU\footnote{\url{http://casu.ast.cam.ac.uk/surveys-projects/vista/technical/data-processing}}). The CASU stacks consist of 16 science images, one per detector, and their corresponding confidence map. A part of detector 16 has unstable gain, and we therefore did not use data from this detector. The data reduction performed by CASU includes dark subtraction, flat-fielding, sky subtraction, gain normalization, nonlinearity correction, and astrometric and photometric calibration. 

\begin{figure}
  \centering
  \begin{subfigure}
    \centering
    \includegraphics[width=0.8\linewidth]{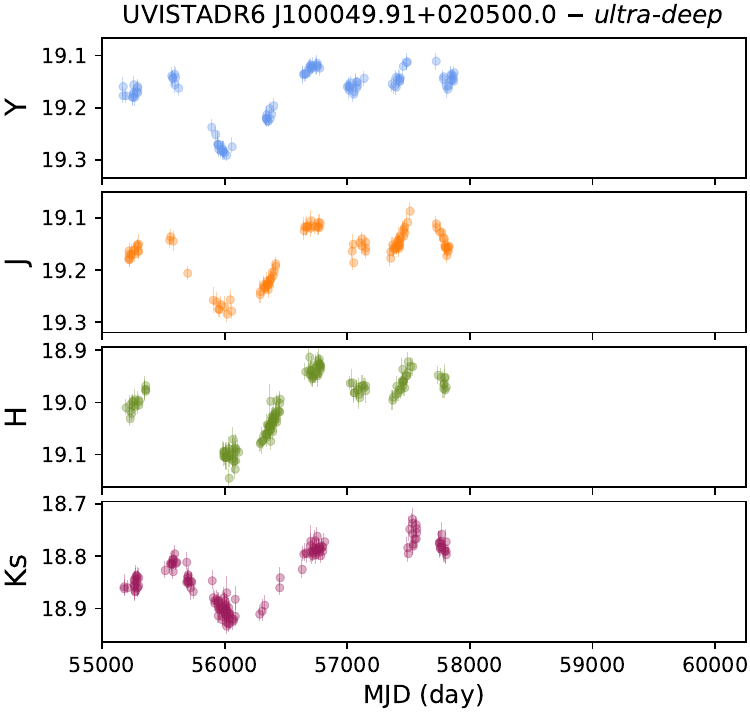}
  \end{subfigure}
  \begin{subfigure}
    \centering
    \includegraphics[width=0.8\linewidth]{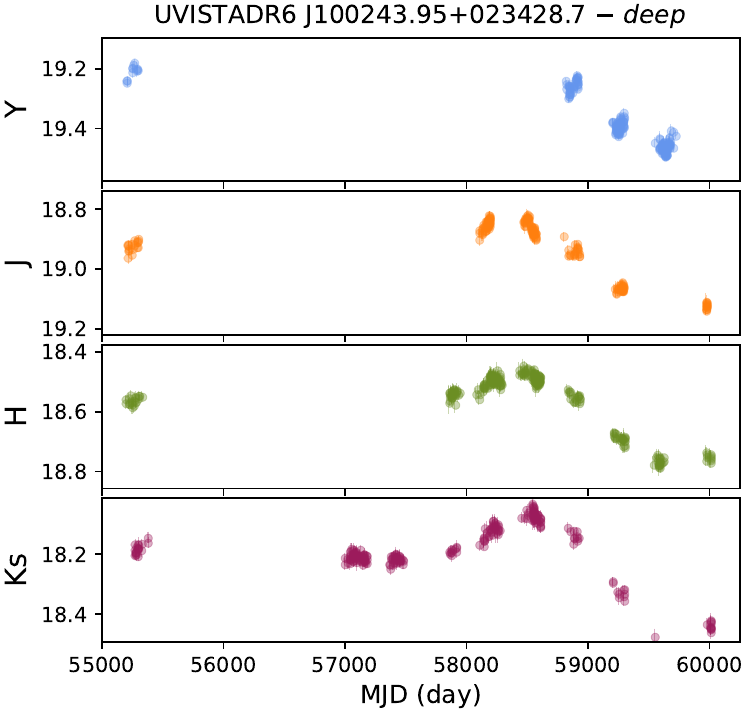}
  \end{subfigure}
  \caption{
NIR light curves of two representative type 1 AGNs, illustrating the observing cadence in the $Y$, $J$, $H$, and $K_s$ bands for the ultra-deep (top) and deep (bottom) stripes.
}
  \label{fig:lc_cadence}
\end{figure}

The temporal sampling mainly depends on the stripe, as the observing time of the UltraVISTA survey was initially spent on the ultra-deep stripes and later extended to the deep ones. In addition, differences in the cadence are present among the bands within each stripe. Figure \ref{fig:lc_cadence} shows two examples of type 1 AGNs light curves constructed following the strategy described in Sect. \ref{section:light curve construction}, illustrating the observing cadence in the $Y$, $J$, $H$, and $K_s$ bands for both the ultra-deep and deep stripes. Objects in the ultra-deep stripes exhibit a more uniform coverage across all bands, with eight consecutive observation sessions, although there are some gaps in $H$ and $K_s$. In contrast, the light curve from the deep stripes spans a longer temporal baseline but exhibits larger gaps across all bands, with the number of observing sessions decreasing from $K_s$ to $Y$. 

\section{Light-curve construction}\label{section:light curve construction} 

Photometry and calibration of the individual OB stacks were performed following an adapted version of the methodology described by \citet{Sanchez2017}. To minimize spurious variability arising from changes in the point spread function (PSF) across epochs, all OB stacks were homogenized to a common seeing $\sigma_0$ by convolving each image with a Gaussian kernel of width equal to $\sqrt{\sigma_0^2  - \sigma^2}$, where $\sigma_0$ correspond to the worst seeing conditions, and $\sigma$ is the width of each individual image. Source detection and photometry were then carried out with \texttt{SExtractor} v2.28.0 \citep{Bertin1996}, using a circular aperture with a radius of 1\arcsec.

Photometric calibration of each epoch was performed by crossmatching the \texttt{SExtractor} catalogs with the DR6 reference catalog, produced by the UltraVISTA consortium and released via ESO (see the release description document), within a radius of $1\arcsec$. For the reference photometry, we adopted the $2\arcsec$-diameter aperture magnitudes provided in the DR6 catalog for each band. We selected only stellar sources (\texttt{CLASS\_STAR} $\geq 0.9$) with reliable photometry (\texttt{FLAG} = 0), magnitudes fainter than 16, and magnitude uncertainties smaller than 0.3. These objects, hereafter referred to as calibration stars, were used to model the magnitude residuals relative to DR6 as a first-order polynomial of the reference magnitude, $m - m_{\text{DR6}} = \alpha + \beta\, m_{\text{DR6}}$. The calibrated magnitudes were then obtained as $m_{\text{cal}} = m - (\alpha + \beta\, m_{\text{DR6}})$. The $\beta$ term accounts for magnitude-dependent residual trends between individual epochs and the DR6 reference, which can arise from the PSF homogenization process. Typical fitted $\beta$ values are 0.0106, 0.0131, 0.0204, and 0.0230 in the $Y$, $J$, $H$, and $K_s$ bands, respectively. Epochs with fewer than 20 matched stars were excluded.

Because \texttt{SExtractor} assumes Poissonian sky noise, it tends to underestimate photometric uncertainties in the presence of correlated noise, that arises from the stacking and convolution procedures. This background noise is particularly dominant for faint sources. To mitigate this, we adopted a different strategy. We measured the rms of the magnitude residuals, $m - m_{\text{DR6}}$, of the calibration stars in bins of magnitude. A polynomial was then fitted to the relation between $\log(\mathrm{rms})$ and the median magnitude of each bin. This model was subsequently used to assign photometric uncertainties to all sources in our \texttt{SExtractor} catalogs. This empirical fit to the scatter as a function of magnitude encapsulates all sources of error, including background noise; therefore, this contribution was not added separately, as was done in \citet{Sanchez2017}. This approach improves the noise estimation for faint sources, where background noise dominates the uncertainty. Furthermore, it refines the method of \citet{Sanchez2017}, who adopted an average rms of the residuals ($m - m_{\text{DR6}}$) across all magnitudes, leading to overestimated errors for bright sources.

The light curves were constructed by crossmatching the calibrated catalogs with the DR6 reference catalog using a 1\arcsec\ matching radius. Epochs with \texttt{FLAG} $\neq 0$, photometric errors exceeding twice the mean magnitude uncertainty, or deviations larger than $5\sigma$, where $\sigma$ is the standard deviation of the light curve, were excluded. We also discarded epochs for which the total rms value exceeds the 95th percentile of the calibration residuals ($m - m_{\text{DR6}}$) distribution, as these epochs typically correspond to poor–quality images. Additionally, light curves with fewer than 20 valid epochs in any individual NIR band after cleaning were discarded. Finally, to minimize edge effects, we excluded a margin of 0.03 deg in both right ascension and declination from all detectors chips, as photometric errors near the borders are unreliable due to the lower effective exposure time. After cleaning, a total of 68,079 sources with light curves in all four NIR bands were retained for further analysis; we refer to this set as the ``clean light-curve sample.''

\begin{figure*}
    \centering
    \includegraphics[width=\textwidth]{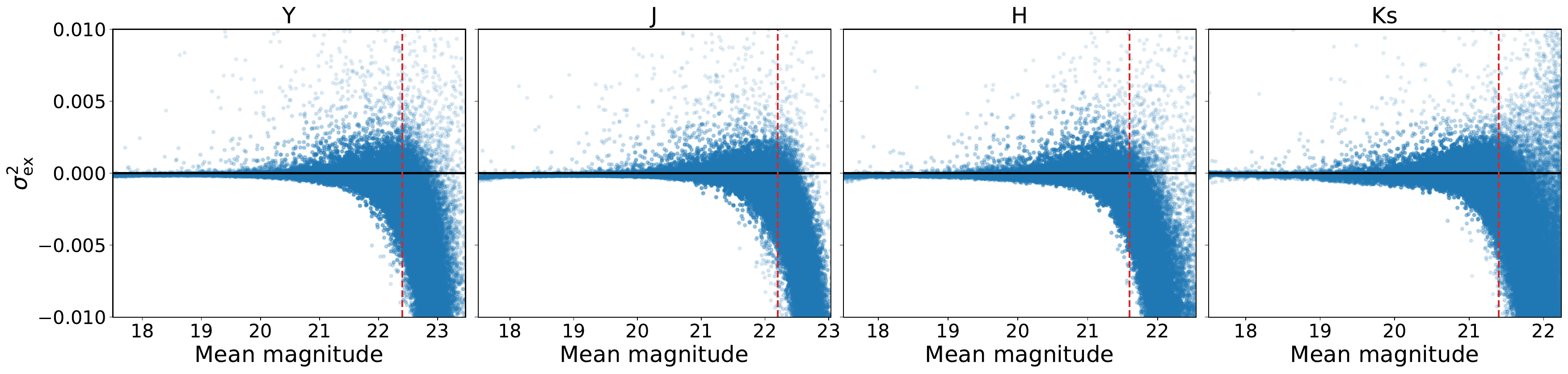}
    \caption{
Distribution of excess variance as a function of mean magnitude for all sources in the clean light-curve sample. Red vertical dashed-lines indicate the adopted magnitude cuts: $22.4$, $22.2$, $21.6$, and $21.4$ mag in the $Y$, $J$, $H$, and $K_s$ bands, respectively.
}
    \label{fig:exvar_mag}
\end{figure*}

We computed the excess variance ($\sigma_{\text{ex}}^2$; described in Sect. \ref{section:features}) separately for each object in each of the four NIR bands. This feature quantifies the intrinsic variability amplitude by comparing the observed scatter of the light curve with that expected from the photometric errors. Figure \ref{fig:exvar_mag} shows the distribution of excess variance as a function of the mean magnitude for each band. The bulk of the distribution shown in Fig. \ref{fig:exvar_mag} corresponds to the photometric errors of nonvariable sources, but the cloud of positive excess variance corresponds to real variability which will be examined further in Sect. \ref{section:results}. At fainter magnitudes, the excess variance systematically drops to negative values, indicating that the photometric uncertainties are overestimated. This behavior is driven by the lack of sufficient calibration stars at the faint end, which prevents a reliable determination of the rms of the magnitude residuals beyond these limits. As a consequence, the polynomial fit used to model the photometric errors becomes poorly constrained at the faintest magnitudes, leading to unreliable error estimates. To mitigate these effects, we applied magnitude cuts of $22.4$, $22.2$, $21.6$, and $21.4$ mag in the $Y$, $J$, $H$, and $K_s$ bands, respectively, which correspond to the faintest calibration stars available in each band. These limits ensure that the excess variance distribution remains approximately centered around zero for nonvariable sources. We also excluded sources brighter than $17.5$ mag in all bands, as AGNs at these magnitudes are scarcely represented in the catalogs used to construct our AGN sample. After applying these selections, the final dataset comprises 26,627 sources with reliable multiband light curves, which we refer to as our ``valid light-curve sample.'' 

\begin{figure}
    \centering
    \includegraphics[width=0.45\textwidth]{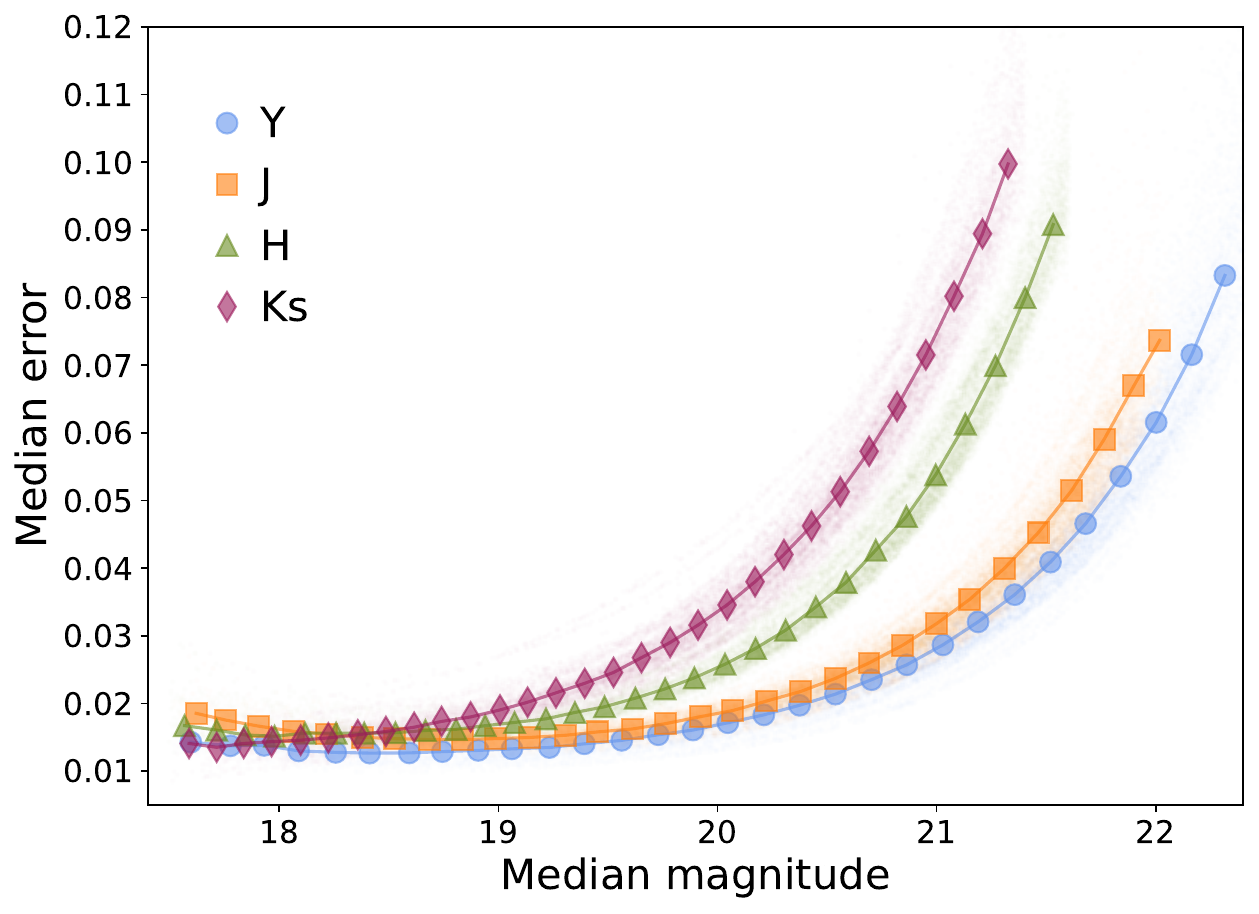}
    \caption{Median photometric error as a function of source median magnitude for the NIR light curves in the valid light-curve sample. Binned medians are shown for each band, with circles, squares, triangles, and diamonds representing the $Y$, $J$, $H$, and $K_s$ bands, respectively.}
    \label{fig:errvsmag}
\end{figure}

Figure \ref{fig:errvsmag} shows the median photometric error of the NIR light curves in our valid light-curve sample as a function of the median magnitude. We observe that for fainter sources the photometric uncertainties are larger in the redder bands, while for brighter objects the errors tend to be comparable across all bands. This trend is likely related to the brighter sky background at longer wavelengths and to the shorter effective exposure times in the redder bands, which on average include $0.5$ h exposures compared to the $1$ h in the bluer bands.

\section{AGN sample}\label{section:agn sample}

The COSMOS field is one of the most studied extragalactic survey areas, with a large number of sources already classified as AGNs, based on different diagnostics. In this work, we constructed our AGN sample using three complementary samples: X-ray AGNs, selected from the \textit{\textit{Chandra}} COSMOS-Legacy Survey \citep{Marchesi2016} and the \textit{XMM-Newton} survey of the COSMOS field \citep{Brusa2010}, with optical spectroscopy counterpart, the Million Quasars catalog (Milliquas; \citeauthor{Flesch2023} \citeyear{Flesch2023}), and the AGN/QSO Value-Added Catalog\footnote{\url{https://data.desi.lbl.gov/public/dr1/vac/dr1/agnqso} } (Juneau et al. in prep) from the First Data Release of the Dark Energy Spectroscopic Instrument (DESI; \citealt{DESIDR1}).

From the \citet{Marchesi2016} catalog, both broad line AGNs (BLAGNs), i.e., sources with at least one broad emission line (FWHM > 2000 km\,s$^{-1}$) in their optical spectra, and non-BLAGN sources, which include narrow line AGNs (NLAGNs) and possibly star-forming galaxies, are considered. From the \citet{Brusa2010} catalog, sources classified as BLAGNs and NLAGNs are also included. In the \textit{XMM-Newton} survey the spectroscopic classification follows the same criteria as in the \textit{Chandra} survey, but NLAGNs have a rest-frame hard X-ray luminosity $> 2 \times 10^{42}$ erg\,s$^{-1}$ to discriminate between star formation processes and accretion processes. For the \textit{Chandra} sample, in addition to the spectroscopic classification, we can recover the X-ray classification into obscured and unobscured sources following \citet{Sanchez2017}. Sources with $0 \leq N_{\rm H} < 10^{22}\,\mathrm{cm}^{-2}$ are classified as unobscured (XR I), while those with $N_{\rm H} \geq 10^{22}\,\mathrm{cm}^{-2}$ are classified as obscured (XR II). For the \textit{XMM-Newton} sample, the classification relies on the hardness ratio,  $\text{HR} = (\text{H}- \text{S})/(\text{H}+ \text{S})$, here H are the hard-band counts and S the soft-band counts. Sources with $\mathrm{HR} > 0.2$ are considered obscured and the rest unobscured. We adopt the \textit{Chandra}-based classification whenever available, as it is derived from the $N_{\rm H}$ and is less affected by redshift than the HR. Otherwise, we use the \textit{XMM-Newton} classification. The combined X-ray sample contains 1780 AGNs, of which 645 have counterparts in the valid light-curve sample. 

To complement the X-ray sample, we crossmatched our valid light-curve sample with the latest version (v8) of the Milliquas catalog \citep{Flesch2023}, which compiles all published type 1 AGN/sQSOs, and type 2 AGNs up to 30 June 2023. In this catalog, sources are classified as QSOs (broad-line, core-dominated), AGNs (Seyfert type 1, host-dominated), NLQSO (narrow-line, core-dominated), and NLAGNs (Seyfert type 2, host-dominated). A total of 561 sources from this catalog have counterparts in our valid light-curve sample.
 
We further included the DESI AGN/QSO Value-Added Catalog (Juneau et al. in prep). We identified type 1 AGNs according to the following criteria: (i) classification as QSO by the \textit{Redrock} pipeline through spectral template fitting; (ii) broad-line detection by the Mg \textsc{II} afterburner; (iii) reclassification as QSO by the machine-learning algorithm QuasarNet; or (iv) the presence of broad emission lines (FWHM $\geq 1200$ km\,s$^{-1}$) in H$\alpha$, H$\beta$, Mg II, and/or C IV. Type 2 AGNs are defined as sources lacking broad emission lines and exhibiting at least one classical BPT diagnostic \citep{BPT1981, Veilleux1987} consistent with a Seyfert classification. These diagnostics include the [N\,\textsc{ii}] diagram \citep{Kewley2001, Kauffmann2003, Schwinski2007}, the [S\,\textsc{ii}] diagram, and the [O\,\textsc{i}] diagram \citep{Kewley2006, Law2021}. In addition, we adopted the ``blue diagram'' \citep{Lamareille2004, Lamareille2010}, which relies on [O\,\textsc{ii}], [O\,\textsc{iii}], and H$\beta$ emission lines, allowing the classification of emission-line galaxies up to $z \sim 1$. From this diagnostic, we selected only those galaxies consistent with a Seyfert 2 classification. A total of 450 DESI sources have counterparts in our valid light-curve sample.

\begin{figure}
\centering
 \begin{subfigure}
    \centering
    \includegraphics[width=0.55\linewidth]{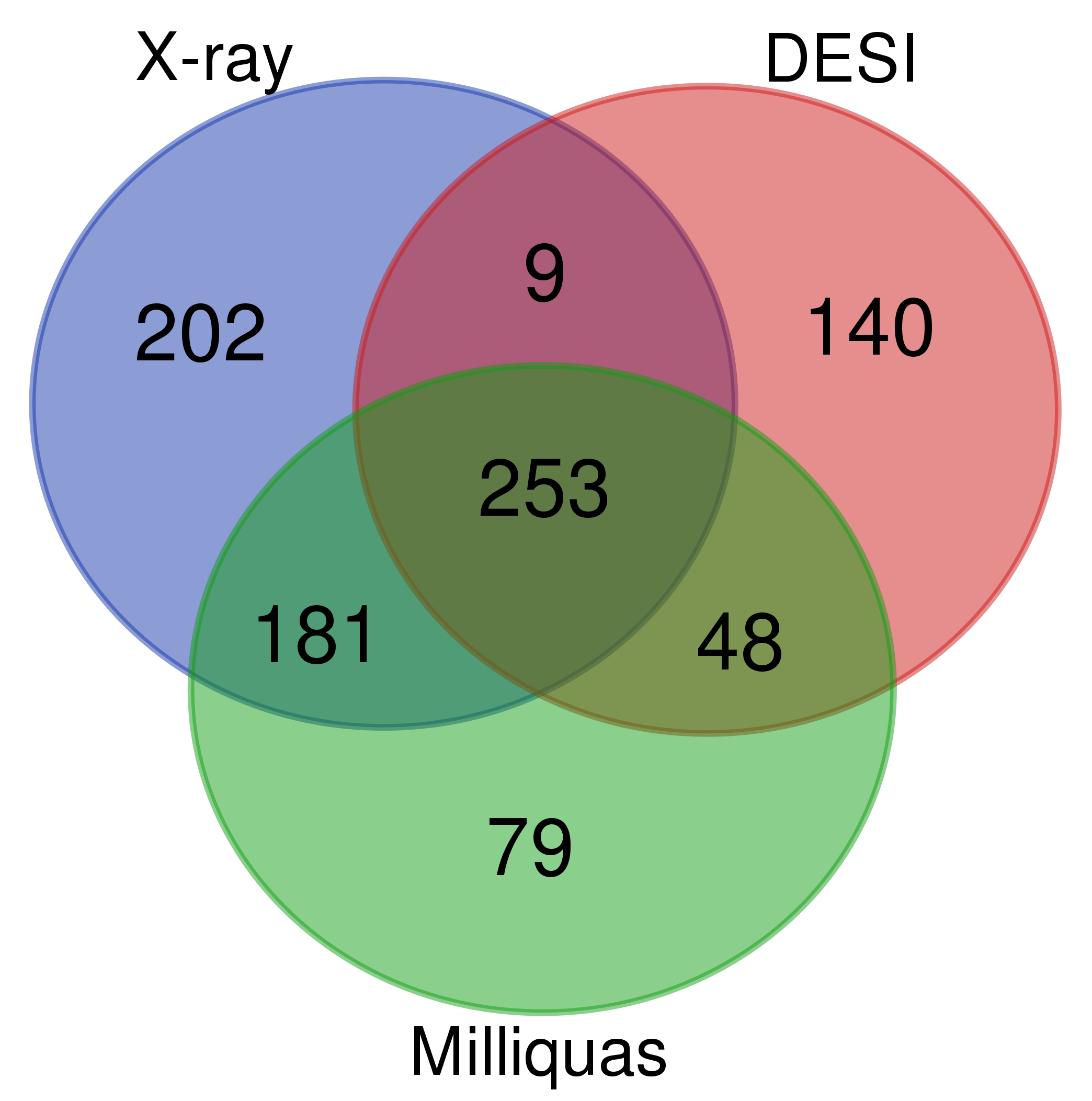}
  \end{subfigure}
  \begin{subfigure}
    \centering
    \includegraphics[width=\linewidth]{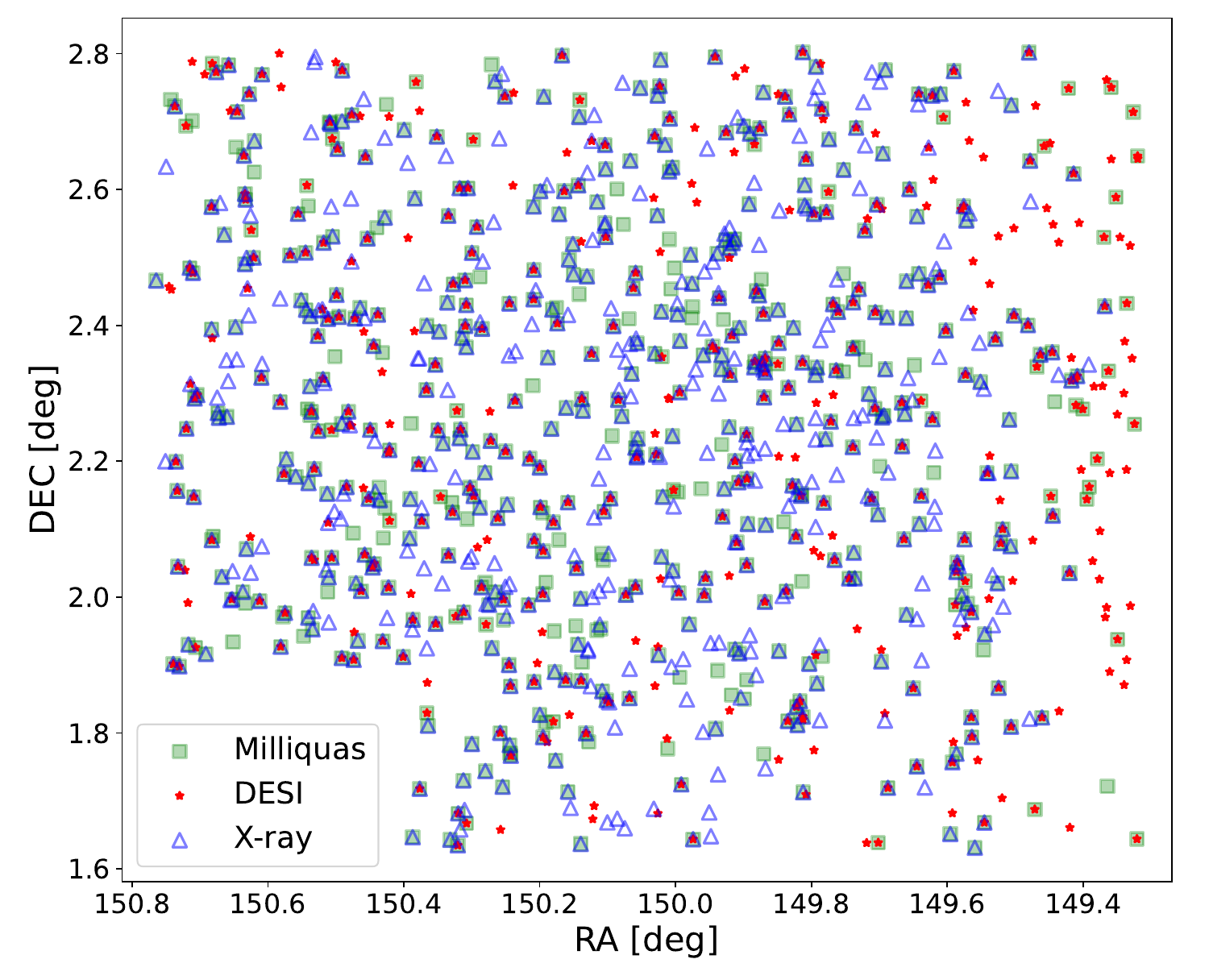}
  \end{subfigure}
\caption{
Venn diagram (top) illustrating the overlap among the three AGN samples used in this work: X-ray (blue), DESI (red), and Milliquas (green).  
The bottom panel shows the spatial distribution of the combined AGN sample across the UltraVISTA field. Squares, stars, and triangles represent Milliquas, DESI, and X-ray sources, respectively.}
\label{fig:venn}
\end{figure}

The top of Fig. \ref{fig:venn} shows the Venn diagram illustrating the overlap between the three AGN samples used in this work: X-ray, DESI, and Milliquas sources. The combined sample comprises a total of 912 AGNs. From this point onward, we refer to this dataset as our AGN sample. The bottom panel of Fig. \ref{fig:venn} displays the spatial distribution of the AGN sample across the UltraVISTA field. The \textit{Chandra} and \textit{XMM-Newton} surveys do not fully cover the UltraVISTA field, leaving the eastern region outside their footprint. In this area, most detected sources correspond exclusively to DESI or Milliquas AGNs. Across the remainder of the UltraVISTA field, several AGNs identified in DESI and Milliquas also lack X-ray counterparts, unveiling a population of AGNs in the COSMOS field that although have X-ray coverage do not have X-ray counterparts. More than half of these sources are classified as type 2 AGNs.

Because several objects are present in more than one catalog, discrepancies in their classification can arise. In this work, we define type 1 AGNs as sources classified as BLAGNs in the X-ray catalogs, as type 1 in DESI, and as QSO/AGN in Milliquas. Similarly, type 2 AGNs are those classified as non-BLAGNs/NLAGNs in the X-ray catalogs, as type 2 in DESI, and as NLQSOs/NLAGNs in Milliquas. To construct a consistent AGN sample, we first identified the sources with a unique and consistent classification across all catalogs, yielding 344 type 1 and 494 type 2 AGNs.

For the remaining 74 objects with conflicting classifications, we applied a hierarchical reassignment scheme. We first adopted the DESI classification, which we consider the most reliable owing to its spectral coverage from the UV to NIR (3600–9800 \AA) and its moderate spectral resolution ($R \sim 2000$–5000; \citealt{DESI2022}), along with the use of additional afterburner algorithms for improved QSO identification \citep{Chaussidon2023, Alexander2023}. When DESI classifications were not available, we adopted the spectroscopic classifications from the \textit{Chandra} and \textit{XMM-Newton }catalogs, which are primarily based on Magellan/IMACS \citep{Trump2007,Trump2009}, VIMOS/zCOSMOS \citep{Lilly2007,Lilly2009}, SDSS and DEIMOS spectroscopy \citep{Hasinger2018}. Finally, we adopted the classifications from the Milliquas catalog, which compiles heterogeneous diagnostics from multiple surveys. Although its classifications generally agree with those from DESI and X-ray catalogs, a fraction of its entries are based on nonspectroscopic criteria, which are less reliable for our purposes. 

\begin{figure}
  \centering
  \begin{subfigure}
    \centering
    \includegraphics[width=\linewidth]{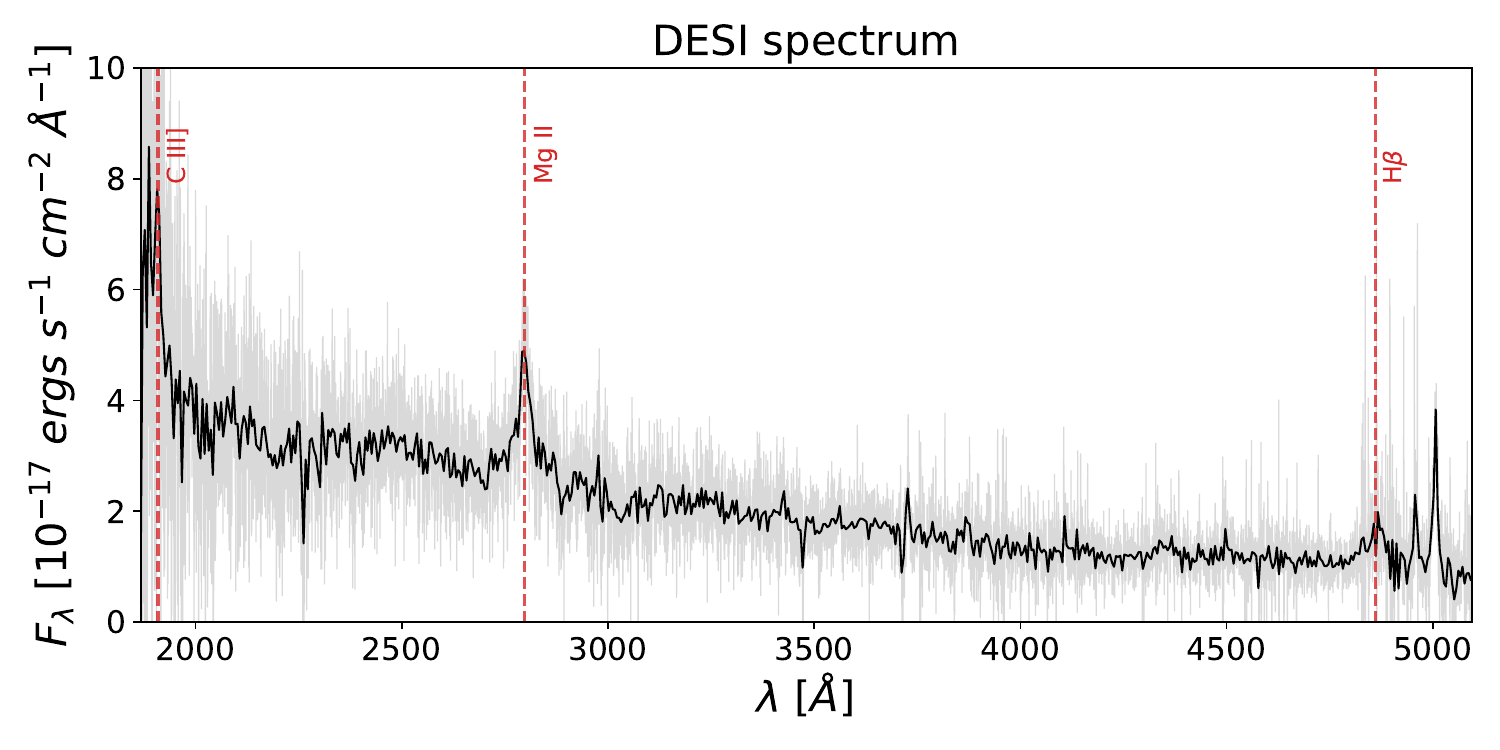}
  \end{subfigure}
  \begin{subfigure}
    \centering
    \includegraphics[width=\linewidth]{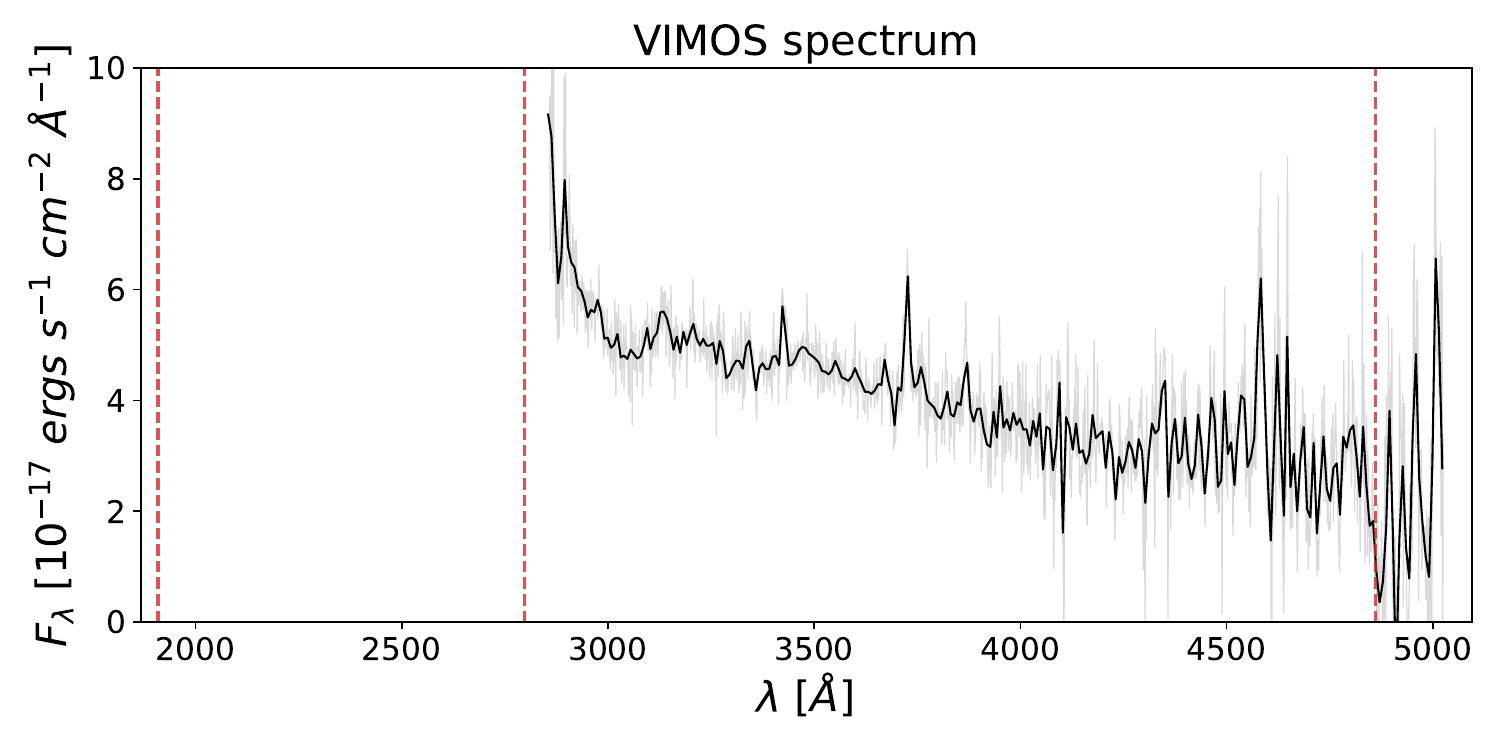}
  \end{subfigure}
\caption{
Comparison of the DESI (top) and VIMOS (bottom) rest-frame spectra of UVISTA–DR6\,J095755.47+022401.1 ($z=0.928$). 
The gray line shows the original spectrum, while the black line displays the spectrum smoothed using the \texttt{FluxConservingResampler} to facilitate the identification of broad emission lines.
Vertical dashed lines indicate the centroids of the broad emission lines.
}
  \label{fig:spectra}
\end{figure}

With the procedure described above, 35 AGNs that were previously classified as type 1 were reclassified as type 2. These objects are likely AGNs whose broad emission lines were not detected in earlier lower-resolution spectra, or whose spectral coverage did not include the broad UV lines. Furthermore, we adopt a slightly stricter definition of type 1 AGNs within the DESI catalog, including sources with broad emission lines of FWHM $\geq 1200$ km\,s$^{-1}$, which is a narrower criterion than that used in the X-ray catalogs. One illustrative example is UVISTA–DR6\,J095755.47+022401.1, which was classified as type 2 in the \textit{XMM-Newton} catalog \citep{Brusa2010} based on its VIMOS spectrum \citep{Lilly2009}, but is now classified as type 1 in DESI DR1. Figure \ref{fig:spectra} shows both rest-frame spectra: in the VIMOS spectrum (lower panel), strong contamination in the red region prevents detection of a broad H$\beta$ component, and the UV broad-line region lies outside the observed wavelength range. In contrast, the DESI spectrum (upper panel) reveals a broad H$\beta$ component and broad UV emission lines. 

The remaining 39 sources were classified as “AGNs” in the Milliquas catalog but as type 2 in DESI or X-ray surveys. These objects were ultimately retained as type 2 AGNs in our classification scheme. After reclassifying all ambiguous cases following this procedure, our final AGN sample comprises 379 type 1 and 533 type 2 sources. The AGN sample selection and classification procedure is summarized in a flowchart presented in Appendix \ref{appendix:agn_sample}

\section{Variability features}\label{section:features}

We characterized the variability of our AGN sample using a set of features that quantify the amplitude and temporal structure of the light curves, as detailed below.

\subsection{Excess variance and $P_{\mathrm{var}}$}

 The normalized excess variance $\sigma_{\text{ex}}^2$ \citep[and references therein]{Allevato2013, Cartier2015, Sanchez2017} quantifies the intrinsic variability amplitude of a source, and it is defined as:
 $(\sigma_{\rm LC}^2 - \bar\sigma_{\text{err}}^2)/\bar{f}$ where $\bar{f}$ is the mean flux, $\sigma_{\rm LC}^2$ the standard deviation of the light curve and $\bar\sigma_{\text{err},i}$ is the mean flux error. The $P_{\mathrm{var}}$ \citep[and references therein]{McLaughlin1996, Cartier2015, Sanchez2017} quantifies the probability that the observed flux variations in a light curve are intrinsic rather than produced by photometric noise, based on a $\chi^2$ test of the flux deviations from the mean. Values of $P_{\mathrm{var}}$ close to 1 indicate statistically significant variability, while lower values suggest consistency with measurement uncertainties.

\subsection{Damped random walk}

Active galactic nucleus light curves are commonly modeled as stochastic processes following a DRW \citep{Kelly2009}, a widely adopted model, although the true PSD of AGN variability can deviate from the DRW form \citep{Arevalo2024}. The DRW
is the simplest model of a family of CARMA models that is characterized by the covariance matrix of the signal
\begin{equation}
   k(t_{ij}) = \sigma_{\rm DRW}^2 \exp(-t_{ij}/\tau_{\rm DRW}),
\end{equation}
where $t_{ij}$ is the time interval between the $i^{\mathrm{th}}$ and $j^{\mathrm{th}}$ epochs, $\tau_{\rm DRW}$ is the relaxation timescale or the characteristic timescale where the PSD breaks, and $\sigma_{\rm DRW}$ is the long-term standard deviation of variability.

In this work, we modeled the AGN light curves using the \texttt{EzTaoX} Python package\footnote{\url{https://github.com/ywx649999311/EzTaoX}} \citep{Yu2026}, which uses celerite \citep{Foreman2017}, a fast Gaussian processes regression library, to compute the likelihood of a set CARMA parameters. Posterior distributions of the DRW parameters, $\tau_{\mathrm{DRW}}$ and $\sigma_{\mathrm{DRW}}$, were obtained through Markov chain Monte Carlo (MCMC) sampling. We adopted the median of each posterior distribution as the best-fit estimate, and the 16th and 84th percentiles as the lower ($\tau_{\mathrm{DRW}}^{\rm le}$, $\sigma_{\mathrm{DRW}}^{\rm le}$) and upper ($\tau_{\mathrm{DRW}}^{\rm ue}$, $\sigma_{\mathrm{DRW}}^{\rm ue}$) $1\sigma$ uncertainties, respectively.

\begin{figure*}
    \centering
    \includegraphics[width=\textwidth]{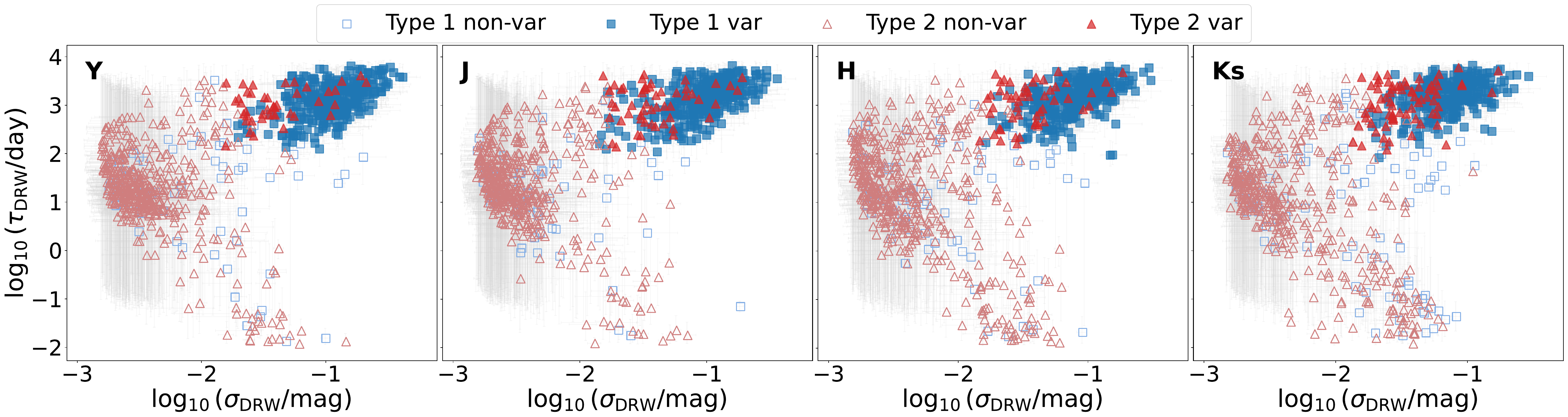}
    \caption{
Distribution of the DRW parameters $\sigma_{\rm DRW}$ and $\tau_{\rm DRW}$ for the AGN sample in the four NIR bands. Type 1 AGNs are displayed as blue squares and type 2 AGNs as red triangles. Filled symbols denote variable sources, while open markers correspond to nonvariable sources. Gray error bars represent the $\pm 1\sigma$ uncertainties.}
    \label{fig:drw}
\end{figure*}

\subsection{Variability criteria}

In \citet{Sanchez2017}, a source is classified as variable if $\sigma_{\text{ex}}^2 - \text{err}(\sigma_{\text{ex}}^2) > 0$ and $P_{\text{var}} \geq 0.95$. However, this criterion is strongly dependent on the photometric uncertainties, which in the NIR are more sensitive to variations in the sky background. As a result, this method can lead to biased variability classifications, particularly for faint sources. To mitigate these effects, we adopt an alternative approach based on the shape of the light curve, using the DRW parameters derived from \texttt{EzTaoX}. Figure \ref{fig:drw} shows the distribution of the DRW parameters, $\sigma_{\rm DRW}$ and $\tau_{\rm DRW}$, for the AGN sample. A clear concentration of type 1 AGNs is seen in the upper-right region of the diagram, corresponding to sources with larger variability amplitudes and characteristic timescales of $\sim$100 to 1000 days. Type 2 AGNs, on the other hand, generally populate the lower-left region of the diagram, exhibiting smaller $\sigma_{\rm DRW}$ and shorter $\tau_{\rm DRW}$ values, as expected if their intrinsic variability is weak or suppressed by obscuration. Furthermore, their inferred $\sigma_{\rm DRW}$ and $\tau_{\rm DRW}$ parameters display large uncertainties, as their light curves provide insufficient signal to constrain the DRW parameters, which cannot therefore be determined with high precision. 

To identify variability consistent with that observed in type 1 AGNs, we require characteristic timescales satisfying $\tau_{\rm DRW} - \tau_{\rm DRW}^{\rm le} > 50$ days. This threshold is consistent with the typical range of quasar variability timescales, which spans from tens to several thousand days \citep[e.g.,][]{Kelly2009, MacLeod2010, Koz2016, Burke2021, Arevalo2024}, and ensures that only sources exhibiting genuine long-term stochastic variability are selected. In addition to the timescale constraint, we impose a minimum variability amplitude requirement. Owing to the photometric quality of the data, intrinsic variations below $\sim$0.01 mag cannot be robustly measured, as this value corresponds to the minimum photometric uncertainty observed for the brightest sources in our sample (see Fig. \ref{fig:errvsmag}). Apparent variability signals below this level are therefore likely to be spurious. To ensure a reliable separation between variable and nonvariable sources, we require $\sigma_{\rm DRW} - \sigma_{\rm DRW}^{\rm le} > 0.01$ mag. 
 
Based on these criteria, we classify as variable those sources that satisfy both conditions on DRW parameters, while those below either threshold are considered nonvariable. As illustrated in Fig. \ref{fig:drw}, variable type 1 AGNs are shown as blue filled squares, while variable type 2 AGNs are represented by red filled triangles. Nonvariable sources are displayed as open markers with the same color and shape. We note that some type 2 AGNs fall in the high-variability region of the type 1s in Fig.  \ref{fig:drw}. The concentration of nonvariable sources at low $\sigma_{\rm DRW}$ values in Fig. \ref{fig:drw} corresponds to the limiting value used in our implementation of the DRW model ($10^{-3}$ mag).

\begin{figure*}
  \centering
\includegraphics[width=\textwidth]{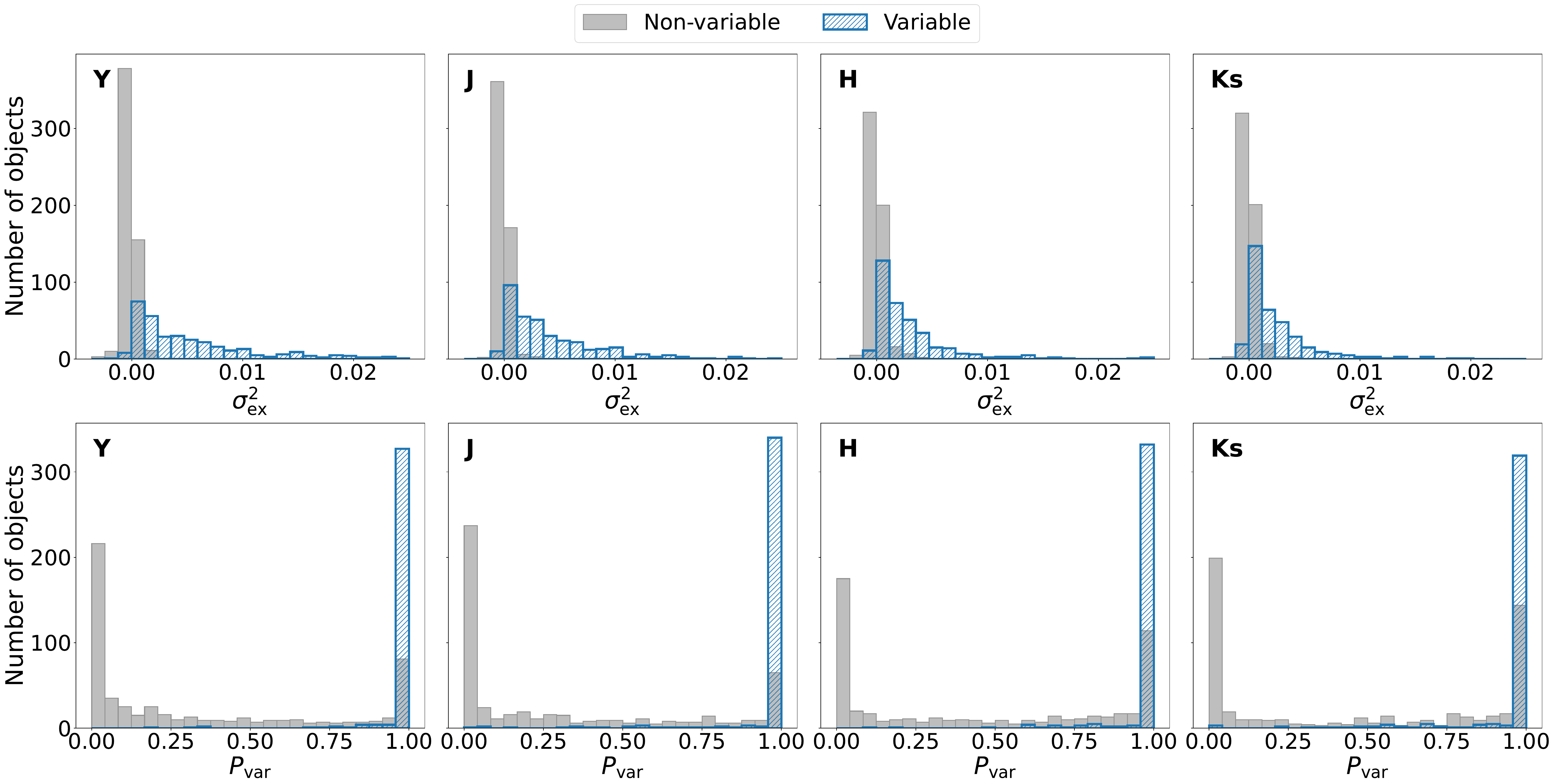}
\caption{
Distributions of the normalized excess variance (top) and the $P_{\rm var}$ (bottom) for AGNs classified as variable (hatched blue) and nonvariable (filled gray) according to the DRW criterion.
}
\label{fig:exvar_pvar}
\end{figure*}

To evaluate the effectiveness of our DRW-based selection, we compare the distributions of the normalized excess variance and $P_{\rm var}$ for sources classified as variable and nonvariable according to the DRW criterion. As shown in Fig. \ref{fig:exvar_pvar}, DRW-selected variables are clearly skewed toward positive $\sigma_{\rm ex}^2$, as expected for variable sources, whereas nonvariables cluster around $\sigma_{\rm ex}^2 \simeq 0$. In addition, $\sim$89-95$\%$ of the DRW-variables sources in the $Y$, $J$, $H$, and $K_s$ bands, exhibit $P_{\rm var}>0.95$, consistent with the standard threshold commonly adopted to identify intrinsically variable objects \citep{Cartier2015,Sanchez2017}. Nonvariable sources, on the other hand, are mostly concentrated at lower probabilities. However, a nonnegligible fraction ($\sim$15-27$\%$) of DRW nonvariable sources also reach $P_{\rm var}>0.95$ values. This effect likely arises from the limited reliability of photometric error estimates in the NIR, which are sensitive to variations in the sky background. Overall, the consistency between the DRW and classical variability criteria confirms that the DRW approach effectively identifies AGN-like variability.

 \subsection{Mexican hat power spectrum (MHPS)}

To estimate the variability amplitude as a function of variation timescale, we will use the Mexican hat filter \citep{Arevalo2012}. This method provides a robust way to characterize variability in light curves with irregular sampling or temporal gaps. The light curve is convolved with two Gaussian filters of slightly different widths, and the difference between the two convolved signals isolates variations around a characteristic timescale. The variance of the filtered light curve represents the normalized variability power at that timescale and, after scaling by the peak frequency of the filter, provides a dimensionless estimate of the variance. In this work, we compute the variances at timescales of 45 and 450 days, which characterize the short- and long-term variability of our AGN sample.

\section{Variability fraction and variability amplitude}\label{section:results}

\begin{table}
\centering
\caption{
Variability fraction of type 1 and type 2 AGNs across the four NIR bands ($YJHK_s$) in the ultra-deep and deep stripes.}
\label{table:pct}
\begin{tabular}{clcr}
\hline
& Filter & ultra-deep (\%) & deep (\%) \\
\hline\hline
\\ [-1.0ex]
Type 1 &\quad $Y$ & $83.6 \pm 2.7$ &  $77.2 \pm 3.1$\\
&\quad $J$ & $84.1 \pm 2.6$ & $82.1 \pm 2.8$ \\
&\quad $H$ & $82.1 \pm 2.8$& $79.9 \pm 3.0$\\
&\quad $K_s$ & $76.4 \pm 3.0$ & $75.5 \pm 3.2$\\ [1.0ex]
\hline
\\ [-1.0ex]
Type 2 &\quad $Y$ & $7.0 \pm 1.6$  & $9.1 \pm 1.7$\\
&\quad $J$ & $7.8 \pm 1.7$& $11.6 \pm 1.9$ \\
&\quad $H$ & $7.4 \pm 1.6$& $12.0 \pm 2.0$\\
&\quad $K_s$ & $10.1 \pm 1.9$ & $16.7 \pm 2.2$ \\ [1.0ex]
\hline\hline
\end{tabular}
\tablefoot{The uncertainties correspond to the $1\sigma$ standard error of the binomial proportion.}
\end{table}

Table \ref{table:pct} summarizes the overall fraction of variable sources in our AGN sample across the four NIR bands, separately for type 1 and type 2 AGNs. To account for differences in cadence and temporal baseline, we present the variability fractions for both the deep and ultra-deep stripes. As expected, type 1 AGNs exhibit a high variability fraction in all NIR bands. Toward redder wavelengths, the variability fraction gradually decreases, consistent with the anticorrelation between the variability amplitude and the emission wavelength \citep{MacLeod2010, Simm2016, Sanchez2017, Li2018, Sanchez2018, Yu2022, Patel2025}. However, in the $Y$ band the fraction is slightly lower than in $J$ mainly in the deep stripes, which can be attributed to the lower cadence of the $Y$-band observations. This reduced sampling limits the ability of the DRW model to recover reliable parameters, resulting in a smaller fraction of detected variables compared to the $J$ band. The ultra-deep stripes, which feature more continuous temporal coverage and a shorter baseline, are more sensitive to short-term variability driven by the accretion disk. Consequently, the variability fractions across all NIR bands tend to be slightly higher than in the deep stripes, where the longer but less regularly sampled light curves may miss a small fraction of rapid variations. In all cases the uncertainties represent the 1$\sigma$ standard error of the binomial proportion, given by $\sqrt{P(1-P)/N}\large$, where $P$ is the variable fraction and $N$ is the total number of sources in each subsample.

Type 2 AGNs exhibit an overall lower variability fraction across all NIR bands. In both observing stripes, the highest variability fraction is found in the $K_s$ band. However, in the deep stripe, the variability fraction increases in all bands compared to the ultra-deep stripe. This behavior likely reflects the longer temporal baseline of the deep observations, which enhances sensitivity to variability on long timescales. The systematically higher fraction of variable sources in $K_s$, observed in both stripes, suggests that this trend is not primarily driven by cadence or temporal coverage, but rather by the intrinsic origin of the emission.

\begin{figure*}
    \centering
    \includegraphics[width=\textwidth]{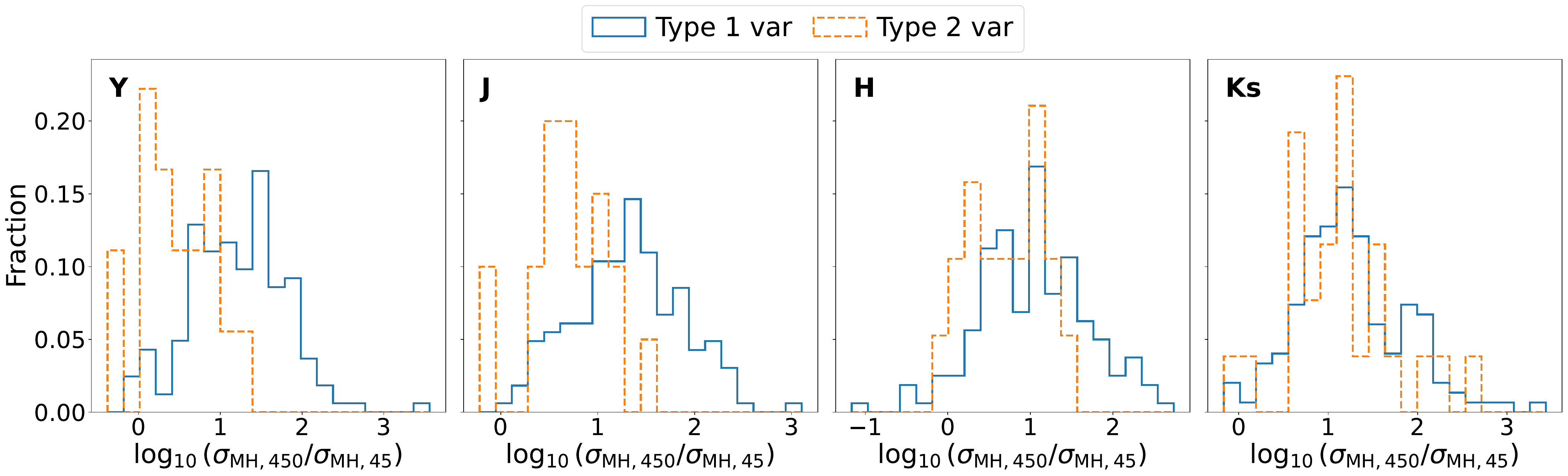}
   \caption{
Distribution of the logarithmic ratio of the MHPS variances measured at intrinsic timescales of 450 and 45 days, $\log_{10}(\sigma_{\rm MH,450} / \sigma_{\rm MH,45})$, for type 1 and type 2 variable AGNs in the ultra-deep stripes across the $Y$, $J$, $H$, and $K_s$ bands. 
}
    \label{fig:mhps_450_45}
\end{figure*}

To further investigate the variability properties of our AGN sample, we compare the MHPS variances computed at intrinsic timescales of 450 and 45 days. This analysis is restricted to the variable sources in the ultra-deep stripes, as these light curves provide the most homogeneous temporal coverage and comparable sampling across all NIR bands. Figure \ref{fig:mhps_450_45} shows the distributions of the ratio between the variances measured at 450 and 45 days, which serves as a proxy for the slope of the PSD.  For type 2 AGNs, we observe a clear evolution of this ratio from the $Y$ to the $K_s$ band, with progressively larger values at redder wavelengths. This trend indicates that the PSD becomes steeper toward $K_s$, implying a larger contrast between long- and short-timescale variability. Such behavior is expected if the $K_s$-band variability includes a stronger contribution from reprocessed emission, which predominantly enhances variability on longer timescales. In contrast, type 1 AGNs do not show a comparable band-dependent evolution of the variance ratio, indicating that the trend observed in type 2 AGNs is not driven by the light-curve sampling.

\section{Discussion}\label{section:discussion}

In this section we discuss the implications of the observed NIR variability for understanding the physical origin of AGN variability and the nature of the population of variable type 2 AGNs.

\subsection{Comparison with \cite{Sanchez2017}}

In \citet{Sanchez2017}, the analysis was based on UltraVISTA DR3, which comprised five years of observations within the ultra-deep stripes across all NIR bands, using counterparts from the \textit{Chandra} catalog. To enable a direct comparison, we repeated the same analysis using our eight-year DR6 observations of the ultra-deep stripes and restrict the sample to the \textit{Chandra} counterparts. Within this \textit{Chandra} subset, we find a clear increase in the variability fractions for both type 1 and type 2 AGNs, with values of 95.7\% (112/117), 97.4\% (114/117), 92.3\% (108/117), and 88.0\% (103/117) for type 1 AGNs, and 15.8\% (27/171), 17.0\% (29/171), 16.4\% (28/171), and 19.9\% (34/171) for type 2 AGNs in the $Y$, $J$, $H$, and $K_s$ bands, respectively. This represents a significant improvement over \citet{Sanchez2017}, where the variability fractions for type 1 AGNs decreased sharply toward redder wavelengths: 77.6\% (152/192), 63.5\% (129/203), 30.9\% (64/207), and 35.4\% (79/223) in $Y$, $J$, $H$, and $K_s$. For type 2 AGNs, a similar decline toward the redder bands was also observed: 8.8\% (28/319), 8.5\% (29/343), 3.0\% (12/405), and 4.9\% (19/388), in $Y$, $J$, $H$, and $K_s$, in contrast to our results, where the variability fraction increases toward longer wavelengths. The overall enhancement in both AGN types arises from the variability selection method adopted in this work, which is less dependent on photometric uncertainties and takes advantage of the longer temporal baseline of the DR6 light curves, allowing a more accurate characterization of intrinsic variability across all NIR bands.

\begin{figure}
  \centering
  \begin{subfigure}
    \centering
    \includegraphics[width=\linewidth]{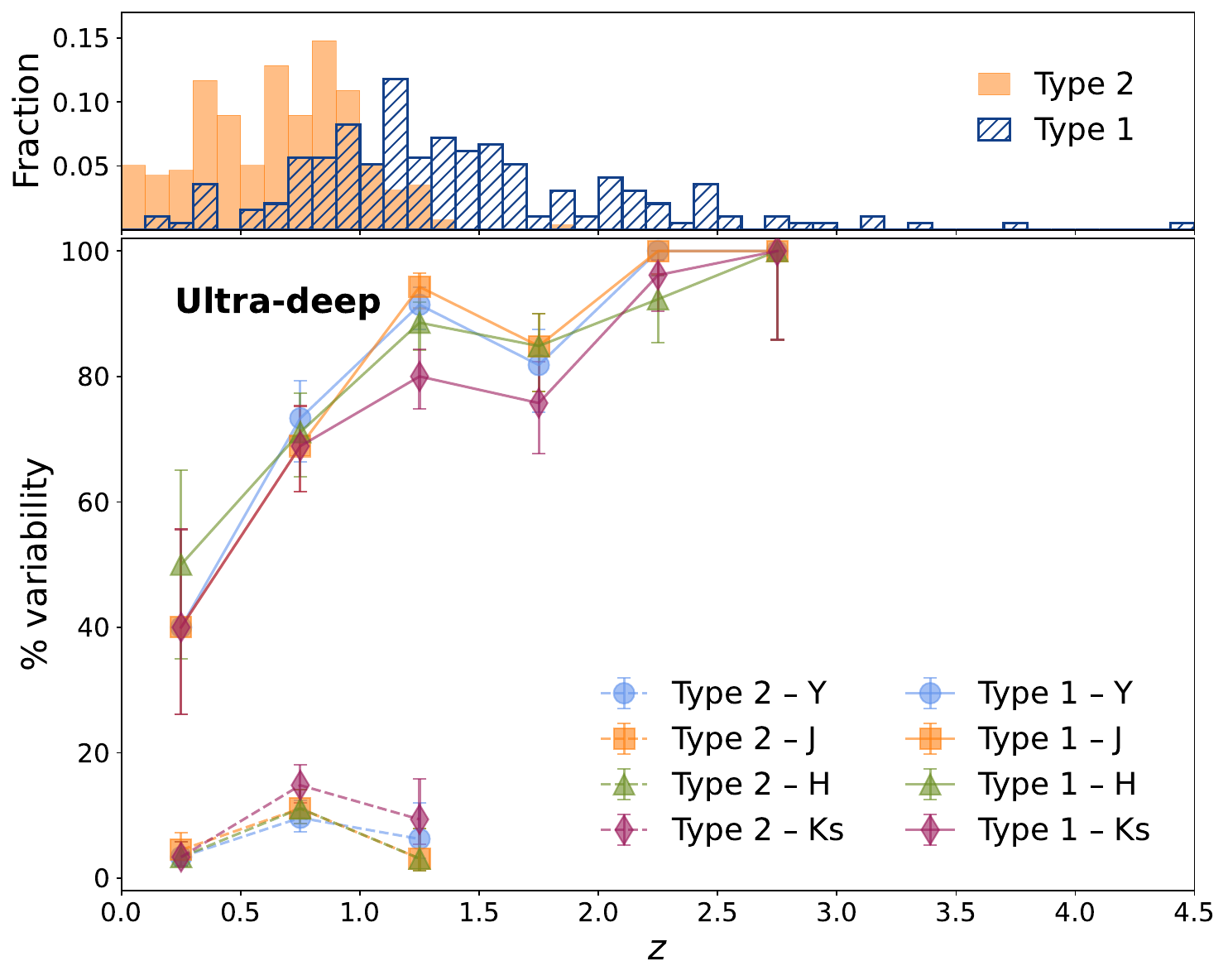}
  \end{subfigure}
  \begin{subfigure}
    \centering
    \includegraphics[width=\linewidth]{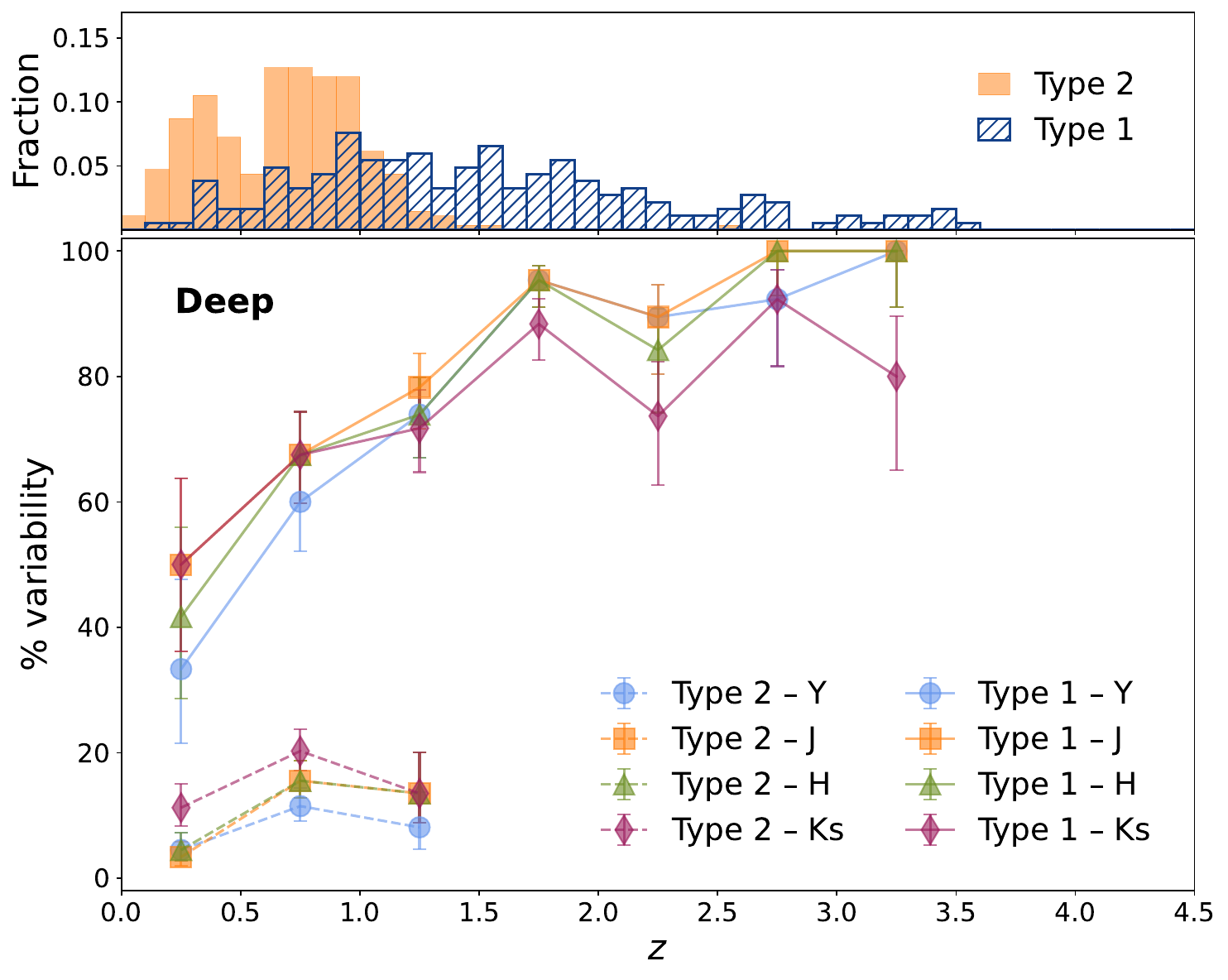}
  \end{subfigure}
\caption{
Fraction of variable sources per redshift bin in each NIR band $YJHK_s$. The top panel corresponds to the ultra-deep stripes and the bottom panel to the deep stripes. In both cases, the upper subpanels show the redshift distribution of type 1 (blue, hatched) and type 2 (orange, filled) AGNs. The lower subpanels display the variability fraction in each band, where different markers indicate the NIR bands (circles: $Y$, squares: $J$, triangles: $H$, diamonds: $K_s$), and solid and dashed lines distinguish type 1 and type 2 AGNs, respectively. Error bars represent $1\sigma$ Wilson confidence intervals, which are asymmetric and more appropriate for binomial proportions, especially in bins with small number statistics.
}
\label{fig:pct_var}
\end{figure}

\begin{figure*}
    \centering
    \includegraphics[width=\textwidth]{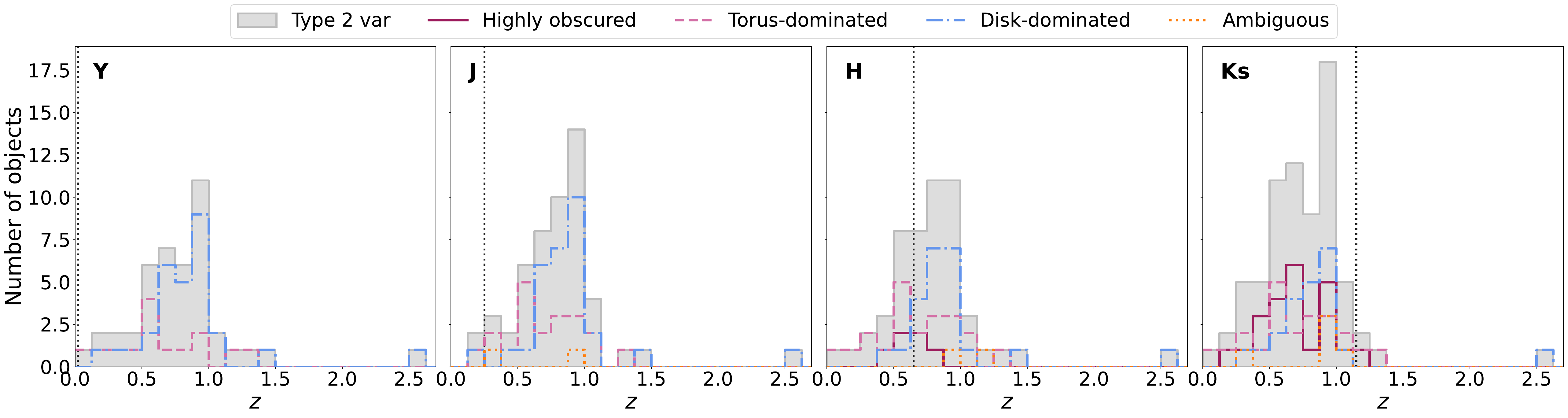}
    \caption{
Redshift distribution of variable type 2 AGNs. 
The gray shaded area represents all variable type 2 AGN sources. 
Solid lines correspond to highly obscured sources, dashed lines to torus-dominated source, dash–dotted lines to disk-dominated sources, and dotted lines to ambiguous cases. Vertical dotted lines mark the redshift at which each observed NIR band corresponds to a rest-frame wavelength of 1 $\mu$m, marking the transition between accretion disk and dusty torus emission.
}
    \label{fig:distribution_agn}
\end{figure*}

\subsection{Variability fraction as a function of redshift and magnitude}

Figure \ref{fig:pct_var} shows the fraction of variable sources for type 1 and type 2 AGNs as a function of redshift, using bins of width $\Delta z = 0.5$, separately for the ultra-deep and deep stripes. For type 1 AGNs, the variability fraction increases with redshift across all NIR bands and remains consistent within the uncertainties between the two observing stripes. At $z>1$, the $K_s$ band generally exhibits slightly lower variability fractions than the bluer NIR bands. This behavior is expected for type 1 AGNs because, at these redshifts, all NIR bands trace accretion–disk emission, where the anticorrelation between variability amplitude and wavelength emission \citep{MacLeod2010, Simm2016, Sanchez2017, Li2018, Sanchez2018, Yu2022, Patel2025} leads to stronger and more easily detectable variability at shorter wavelengths.

Type 2 AGNs exhibit consistently low variability fractions across all NIR bands. In both observing stripes, the highest fractions are found in the $K_s$ band. This behavior arises because, for $z \lesssim 1.15$, the observed $K_s$ band probes rest-frame wavelengths longer than 1 $\mu$m, where the reprocessed emission from the dusty torus dominates. Consequently, variability originating in this extended structure is more readily detected in $K_s$ than in the bluer NIR bands, whose rest-frame wavelengths are increasingly affected by dust extinction, leading to lower fractions of detected variables.

We investigated the behavior of the variability fraction as a function of the observed magnitude in Appendix \ref{appendix:var_mag}, by computing it in bins for each NIR band and for both stripes (see Fig. \ref{fig:pct_var_mag}). For type 1 AGNs, we do not find any significant dependence on magnitude, as the variability fraction remains consistently high across the NIR bands within the uncertainties (see Tables \ref{table:var_mag_udeep} and \ref{table:var_mag_deep}). For type 2 AGNs, the relatively small number of variable sources in each bin leads to large uncertainties, preventing a reliable assessment of any potential trend with magnitude. Given the relatively large number of variable type 1 sources, this suggests that the adopted variability criteria are robust over the full magnitude range probed by our data.

We compared the magnitude and redshift distributions of AGNs in the deep and ultra-deep stripes using two-sample Kolmogorov–Smirnov (K-S) tests. For both type 1 and type 2 AGNs, the magnitude distributions in all bands are statistically consistent between the two stripes, with high $p$-values ($p \gtrsim 0.4$) and small K-S statistics ($D \lesssim 0.15$), indicating no significant differences. Similarly, the redshift distributions do not show statistically significant differences ($p = 0.10$ for type 1 and $p = 0.38$ for type 2). These findings show that the differences in variability fractions between the two stripes cannot be attributed to magnitude or redshift effects, and are instead likely due to the different cadence and temporal baseline of the light curves.

\subsection{Nature of the variable type 2 AGNs}

In total, we identify 43, 52, 52, and 72 type 2 AGNs that exhibit variability in the $Y$, $J$, $H$, and $K_s$ bands, respectively. Overall, 90 of the 533 ($\sim$17$\%$) type 2 AGNs show variability in at least one of these bands. To investigate the physical origin of their variability, we classified the sources into three main groups: disk-dominated, torus-dominated, and highly obscured. Disk-dominated sources show increasing variability amplitude toward shorter wavelengths, torus-dominated AGNs display stronger variations at redder wavelengths, while highly obscured sources exhibit variability only in the reddest bands. A small subset of seven objects present ambiguous variability behavior and cannot be clearly assigned to any of these categories.

\subsubsection{Disk-dominated AGNs}

Disk-dominated AGNs are defined as sources that exhibit variability in the bluer bands ($Y$ and $J$), with variability amplitudes ($\sigma_{\rm DRW}$) in both $Y$ and $J$ larger than those in the any of the redder bands ($H$ and/or $K_s$). We note that variability may also be present in $H$ and $K_s$. We identify 37 such sources. At the redshifts of variable type 2 AGNs, the observed $Y$ and $J$ bands correspond to rest-frame wavelengths shorter than 1 $\mu$m (see Fig. \ref{fig:distribution_agn}). Variability detected in this regime therefore implies partial access to the accretion disk and the BLR. Among the 37 disk-dominated sources, 35 have X-ray counterparts, of which 30 are classified as XR I and 5 as XR II. The high fraction of X-ray unobscured sources supports the interpretation that these AGNs are predominantly viewed through unobscured lines of sight, consistent with a population of weak type 1, where the broad lines are weak and are diluted by host galaxy contamination or by the low signal to noise ratio (S/N), or ``true type 2'' AGNs \citep{Panessa2002, Laor2003, Elitzur2016}, i.e., those that appear to lack a BLR. As shown in Fig. \ref{fig:distribution_agn}, these objects are predominantly located at higher redshifts compared to the overall population of variable type 2 AGNs.

\subsubsection{Torus-dominated AGNs}

Torus-dominated AGNs are sources that show variability in at least one of the redder bands ($H$ and/or $K_s$) and in at least one of the bluer bands ($Y$ and $J$), with larger variability amplitudes in one or both of the redder bands than in both of the bluer bands. We identify 23 such sources in our sample. These AGNs likely represent systems where a faint contribution from the accretion disk becomes directly observable. In this scenario, the variability detected in the bluer bands traces the outer regions of the accretion disk, while the stronger variability observed in $H$ and $K_s$ arises from the more variable inner disk emission reprocessed by the surrounding dusty torus. Because these objects show partial access to the accretion disk, we expect them to correspond primarily to weak type 1 AGNs or ``true type 2'' AGNs. Indeed, 21 of these sources have X-ray counterparts, of which 11 are classified as XR I, consistent with this interpretation. However, the remaining 10 objects are classified as XR II, indicating that a subset of obscured AGNs can still display detectable variability in the bluer NIR bands. In these cases, the variability observed in the $Y$ or $J$ bands likely arises because these wavelengths probe the transition region between the outer accretion disk and the inner edge of the dusty torus. Alternatively, this variability could originate from scattered emission from a compact region close to the nucleus. These torus-dominated AGNs are predominantly found at lower redshifts, as illustrated in Fig. \ref{fig:distribution_agn}.

\subsubsection{Highly obscured AGNs}

Highly obscured AGNs are defined similarly to torus-dominated sources in terms of detected  variability in one or both of the redder bands ($H$ and/or $K_s$). However, in this case, no variability is detected in either of the two bluer bands ($Y$ and $J$). This group comprises 23 objects. The absence of detectable variability at shorter wavelengths indicates that the accretion disk emission is heavily obscured, and that the observed variability in the redder bands is dominated by the dusty torus. As shown in Fig. \ref{fig:distribution_agn}, nearly all variable type 2 AGNs in this category are located at redshifts where the $K_s$ and a fraction of the $H$ band trace rest-frame wavelengths longer than 1 $\mu$m, marking the transition region where the torus contribution becomes dominant. Among these sources, 12 are classified as XR II and 3 as XR I, indicating that most optically obscured AGNs are also X-ray obscured, consistent with expectations from the Unified Model. The small fraction of XR I objects in this group may correspond to weak type 1 AGNs or ``true type 2'' AGNs, whose variability at bluer wavelengths is more strongly affected by host–galaxy and therefore becomes suppressed in the $Y$ and $J$ bands.

To illustrate the characteristic NIR variability signatures of the three groups of variable type 2 AGNs, we present representative examples of their light curves in Appendix \ref{appendix:lc}. Figure \ref{fig:353742} shows an example of a highly obscured AGN classified as XR II at $z = 0.69$, which exhibits clear variability in the $H$ and $K_s$ bands, while the $Y$ and $J$ bands show strongly suppressed variations, as expected for obscured systems. To further investigate its infrared behavior, we include light curves from the unTimely Catalog Explorer \citep{Meisner2023}, a deep time-domain catalog based on the Wide-field Infrared Survey Explorer (WISE) and NeoWISE \citep{Wright2010, Mainzer2014} observations spanning 2010–2020. Pronounced variability is also detected in the mid-infrared (MIR) WISE bands, W1 (3.4 $\mu$m) and W2 (4.6 $\mu$m), reinforcing the interpretation that the observed variations originate in the dusty torus. Figure \ref{fig:262128} presents an example of a torus-dominated AGN classified as XR II at $z = 0.55$, which exhibits variability across all NIR bands, with amplitudes increasing toward redder wavelengths. Consistent with this behavior, strong variability is also observed in the WISE bands. Finally, Fig. \ref{fig:114913} shows a representative disk-dominated XR I AGN, where the variability amplitude is larger in the bluer $Y$ and $J$ bands than in the $H$ and $K_s$ bands, indicating that the variability is primarily driven by the accretion disk.

\begin{figure}
    \centering
    \includegraphics[width=0.45\textwidth]{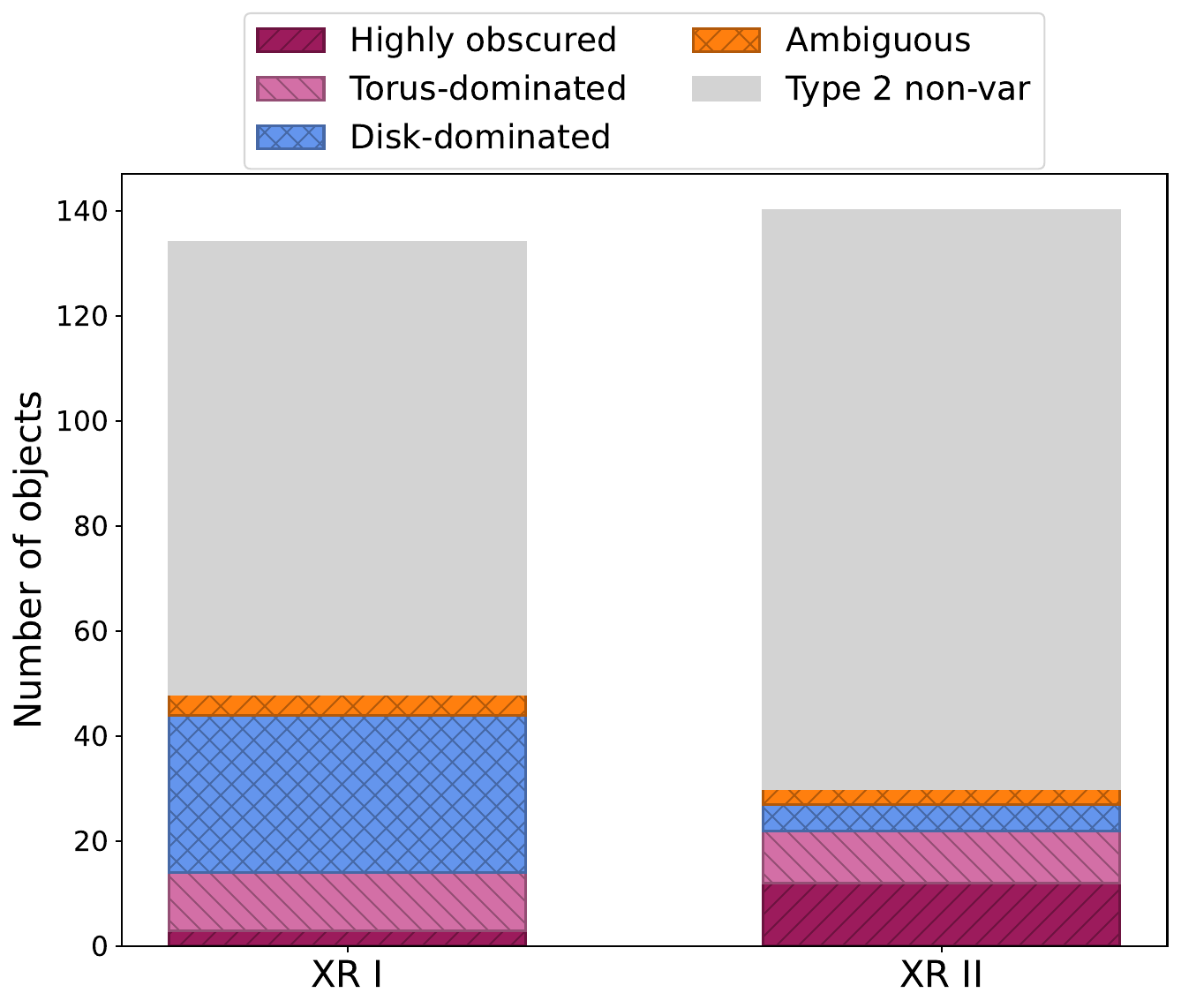}
    \caption{Number of type 2 AGNs according to their X-ray classification. Colored areas represent the number of variable sources in each group.
}
    \label{fig:xray_type2}
\end{figure}

\subsection{X-ray properties of variable type 2 AGNs}

Figure \ref{fig:xray_type2} shows the distribution of all type 2 AGNs according to their X-ray classifications, highlighting the subset identified as variable. In the full type 2 AGN sample, the fractions of XR I and XR II sources are comparable. However, as expected, the majority of variable type 2 AGNs belong to the X-ray unobscured population, primarily within the disk- and torus-dominated groups. As discussed earlier, these objects likely correspond to misclassified weak type 1 AGNs or ``true type 2'' AGNs.

There are 48 variable type 2 AGNs classified as XR I (out of a total of 134 XR I sources), which likely represent candidates for weak type 1 AGNs or ``true type 2'' AGNs. Among them, 35 have available DESI DR1 spectra. We computed the S/N of each spectrum and found that most display low values, typically $\mathrm{S/N} \sim 2$, which prevents a reliable assessment of the presence of broad emission lines. At such low S/N
levels, faint broad components would remain undetectable, making it impossible to determine whether these sources are weak type 1 AGNs or ``true type 2'' AGNs. High-quality spectroscopy with 8–10 m class telescopes will be required to establish their true nature.

We find that 30 of the 140 X-ray obscured sources show detectable variability, mainly among the highly obscured and torus-dominated groups. These objects represent the subset of type 2 AGNs that, according to the Unified Model, are expected to show variability at IR wavelengths. In such systems, the accretion disk is heavily obscured, and the variability we observe arises mainly from the dusty torus. However, this group constitutes only a small fraction of the entire type 2 XR II population. Under the Unified Model, one might expect most type 2 XR II AGNs to display torus-driven variability in the $K_s$ band, yet the vast majority do not. This suggests that, for many type 2 XR II sources, the intrinsic variability of the torus is too weak to detect or is obscured, either by the circumnuclear dust itself or by the host galaxy.

To explore whether some of the nonvariable type 2 XR II sources might exhibit variability at even redder wavelengths, we examined their MIR light curves using W1 and W2 data from WISE and NeoWISE using the unTimely Catalog Explorer \citep{Meisner2023}. These bands probe rest-frame wavelengths deep in the torus-dominated regime, where obscuration effects are reduced. However, MIR light curves are generally less informative for distinguishing obscuration levels in type 2 AGNs, as WISE variability is limited by lower cadence. \citet{Son2023} have shown that the structure function of type 1 and type 2 AGNs in W1 exhibits a similar shape, although type 2 AGNs show less normalization. This suggests that MIR variability provides weaker discriminatory power between obscuration classes. In our sample the WISE data do not reveal a clear separation between variable and nonvariable type 2 XR II sources.

In addition to the sources with an X-ray classification, we identify 259 type 2 AGNs without an XR I/XR II label. Among them, 94 objects do have counterparts in the \textit{Chandra} or \textit{XMM-Newton} catalogs, but lack an X-ray classification because they do not have sufficient counts to obtain reliable constraints on $N_{\rm H}$. Within this subset, 6.4\% exhibit detectable NIR variability. For the remaining 165 type 2 AGNs with no X-ray counterpart, only 3.6\% are classified as variable. This lower fraction reflects the fact that these systems are fainter in X-rays and lie close to or below the detection threshold, making their X-ray identification more challenging. One possibility is that a fraction of these sources corresponds to so-called “switched-off” AGNs \citep{Simmonds2016, Saade2022}, in which the NLR remains visible while the central engine has faded, leading to weak or undetectable X-ray and NIR variability. Alternatively, these objects may be Compton-thick, for which the nuclear emission is strongly suppressed across X-ray and optical/NIR wavelengths, preventing the detection of variability even if the intrinsic activity is still present.

For completeness, we note that the redshift distributions of the X-ray-detected and X-ray nondetected type 2 AGNs are different, with the X-ray-detected sample peaking at higher redshifts. The median redshift of each sample are 0.75 for those with X-ray detections and 0.46 for those without, and a K-S test shows that the distributions are significantly different, with a $p=4.73 \times 10^{-13}$. Their magnitude distributions are also different in the $Y$ and $J$ band, but not in the redder $H$ and $K_s$ bands. The K-S distances and $p$-values, together with the distributions are shown in Appendix \ref{appendix:xray_nonxray}. The lower fraction of variable objects in the X-ray undetected sample coincides with the lower variable fraction of lowest redshift objects, where this sample is concentrated. However, it is not clear which is the primary source of the correlation.

\section{Conclusions}\label{section:conclusions}

In this work, we studied the NIR variability of known AGNs in the COSMOS field using data from UltraVISTA DR6, which provides multiepoch observations spanning 14 years in the $Y$, $J$, $H$, and $K_s$ bands across two observing stripes. The AGN sample was constructed by combining sources from the \textit{Chandra} and \textit{XMM-Newton} X-ray surveys, the DESI AGN/QSO Value-Added Catalog, and the Milliquas catalog, which together provide a comprehensive set of known AGNs in the COSMOS field. Our main conclusions are summarized as follows:

\begin{itemize}
    \item The DRW method proves to be a robust and reliable approach for identifying AGN-like variability in the NIR. Compared to previous approaches such as \citet{Sanchez2017}, which identified variable sources using only two variability statistics derived from the variance in the light curves and the errors (namely the excess variance $\sigma_{\rm{ex}}^2$ and $P_{\rm var}$), our method improves the variability selection by being less sensitive to the photometric errors, which yields a more consistent and homogeneous classification across the NIR bands.
    \item Type 1 AGNs display a high fraction of variable sources in all NIR bands, as expected from the fact we have an unobscured view of the accretion disk. The variability fraction decreases toward redder wavelengths, which is consistent with the know anticorrelation between the variability amplitude and the emission wavelength.
    \item The observed fraction of type 1 variable AGNs increases with redshift, which reflects the fact that at higher redshift the observed NIR bands trace bluer rest-frame wavelengths dominated by the accretion disk, where the variability amplitude is intrinsically higher and therefore easier to detect.
    \item Between $\sim$7-17\% of type 2 AGNs are found to be variable. The variability fraction rises toward the $K_s$ band, reflecting the growing contribution of the dusty torus at redder wavelengths, where its reprocessed emission becomes the dominant source of variability.
    \item Based on the wavelength dependence of the variability amplitude, we classified the variable type 2 AGNs into three groups: disk-dominated, torus-dominated, and highly obscured. The disk-dominated group represents 41\% of all type 2 variables and consists mainly of X-ray unobscured sources (86\% XR I, 14\% XR II). The torus-dominated group accounts for 25.5\% of the type 2 variables, with a comparable fraction of X-ray unobscured (52\%) and obscured (48\%) AGNs. The highly obscured group comprises 25.5\% of the type 2 variables and is dominated by X-ray obscured sources (80\% XR II, 20\% XR I). 
    \item The X-ray analysis reveals that $\sim$36\% of type 2 XR I sources are variable. These objects likely correspond to misclassified systems, either weak type 1 AGNs or ``true type 2'' AGNs.
    \item We find that about 21\% of XR II (obscured) type 2 AGNs exhibit detectable variability, even though, within the redshift range of our sample, the $K_s$ band probes rest-frame wavelengths where torus emission should contribute significantly. This suggests that the variability of the dusty torus in many XR II AGNs is either too weak to be detected or remains obscured, either by the circumnuclear dust itself or by the host galaxy.
    \item Type 2 AGNs without X-ray counterparts show the smallest fraction (3.6\%) of variable objects. These objects could correspond to very highly obscured, Compton-thick AGNs, where all the nuclear emission is highly suppressed in our line of sight, or, alternatively, to faded AGNs, where only the distant NLR region retains AGN signatures. 
    \item  The observed connection between NIR variability and X-ray obscuration supports the unified model, where the observed differences arise mainly from orientation and the relative contribution of the accretion disk and the torus. The discrepancies between optical and X-ray classifications can be largely attributed to the limited detectability of weak broad emission lines in low-S/N spectra or strong host galaxy contamination. 
    \item NIR variability provides an effective and independent diagnostic to confirm optical classifications. In particular, it allows us to identify weak type 1 AGNs that are misclassified as type 2 AGNs in optical surveys, since their detectable NIR variability reveals partial access to the accretion disk.
\end{itemize}

Overall, our results demonstrate that NIR variability is a powerful diagnostic for probing the central structures of AGNs as it provides a means to separate contributions from the disk and torus. The extended time base and depth of the UltraVISTA DR6 survey enables the robust identification of variable type 2 AGNs. Future follow-up spectroscopy with large-aperture telescopes and high-cadence infrared monitoring will be essential to confirm the physical nature of these sources and to refine our understanding of AGN variability.

Looking ahead, this work opens the possibility of extending our analysis to wider fields. In particular, the VISTA Deep Extragalactic Observations (VIDEO) survey \citep{Jarvis2013}, which provides deep multiepoch NIR imaging over a substantially larger area than UltraVISTA, will allow a direct application of our methodology to a broader AGN population. Furthermore, several ongoing and upcoming IR time-domain surveys, such as the Wide-Field Infrared Transient Explorer (WINTER; \citealt{Lourie2020}, the PRime-focus Infrared Microlensing Experiment \citep{Sumi2025}, SPHEREx \citep{Bock2026}, and the\textit{ Nancy Grace Roman} Space Telescope will play a key role in advancing our understanding of AGN variability at the NIR, and offer new opportunities to probe obscuration and AGN structure.

\begin{acknowledgements}

We acknowledge financial support from ESO Science Internship program (JCM), Millennium Science Initiative Program NCN$2023\_002$ (JCM, PA, PL), Millennium Science Initiative, AIM23-0001 (JCM, PA, PSS), FONDECYT Regular 1241422 (PA, PL) and CAV, CIDI N. 21 U. de Valparaíso, Chile (PA). WY acknowledges support from the Dunlap Institute for Astronomy \& Astrophysics at the University of Toronto. The Cosmic Dawn Center (DAWN) is funded by the Danish National Research Foundation under grant DNRF140. Based on observations collected at the European Southern Observatory under ESO programs 179.A-2005, 198.A-2003, 1104.A-0643 and 110.25A2 and on data obtained from the ESO Science Archive Facility with DOI https://doi.
org/10.18727/archive/52, and on data products produced by CANDIDE and the Cambridge Astronomy Survey Unit on behalf of the UltraVISTA consortium.
PA and JCM are grateful for the ESO hospitality. 
\end{acknowledgements}

\bibliographystyle{aa}
\bibliography{bibliography}

\appendix

\section{AGN sample}\label{appendix:agn_sample}

Schematic overview of the AGN sample selection and classification procedure. The flowchart summarizes the sequence of steps followed to construct the final sample and the adopted classification scheme.

\begin{figure*}
    \centering
    \includegraphics[width=0.8\linewidth]{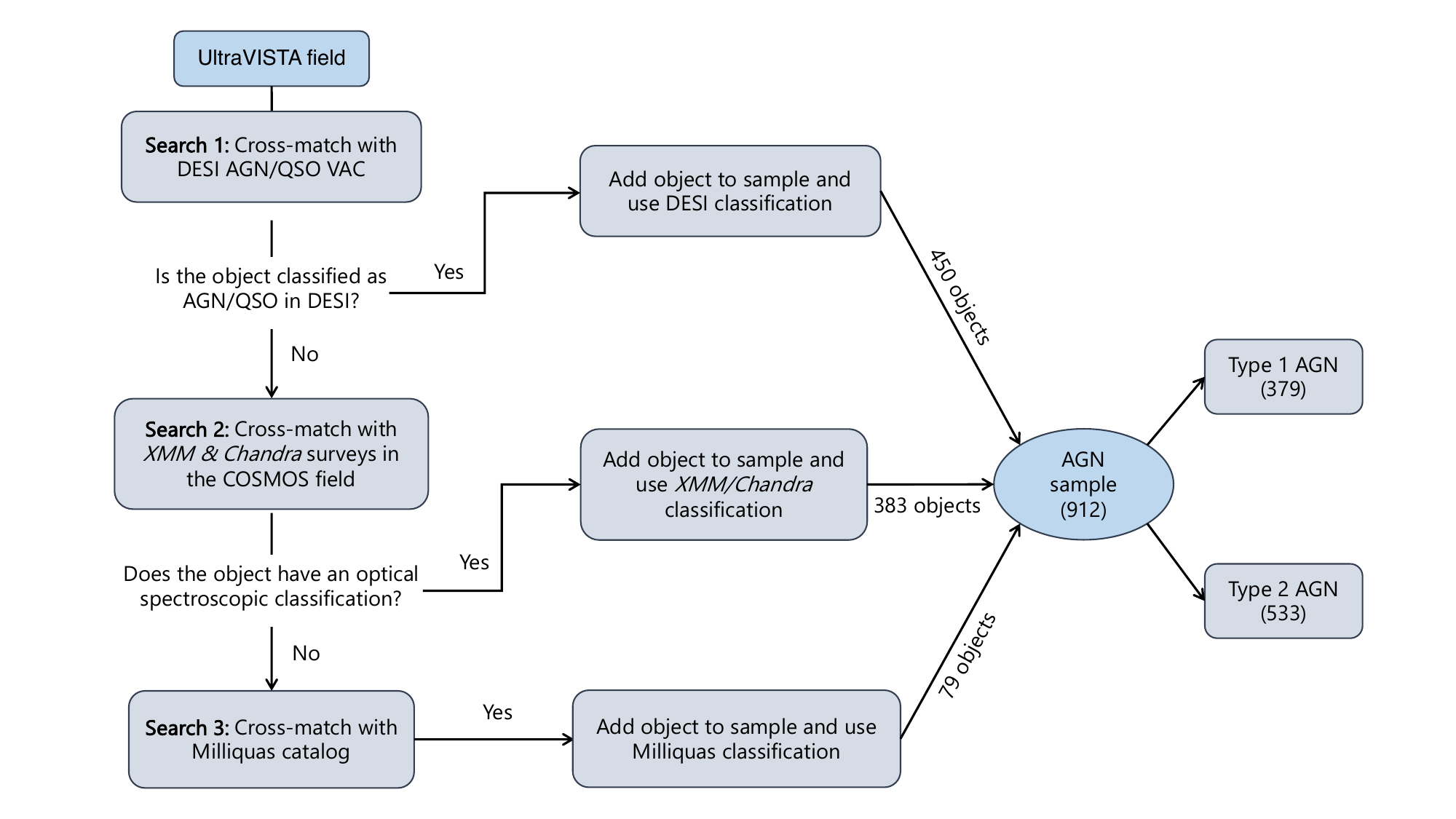}
    \caption{
    Flowchart illustrating the AGN sample selection and classification process.
    }
    \label{fig:flow_chart}
\end{figure*}

\section{Variability fraction as a function of magnitude}\label{appendix:var_mag}

\begin{figure}
  \centering
  \begin{subfigure}
    \centering
    \includegraphics[width=\linewidth]{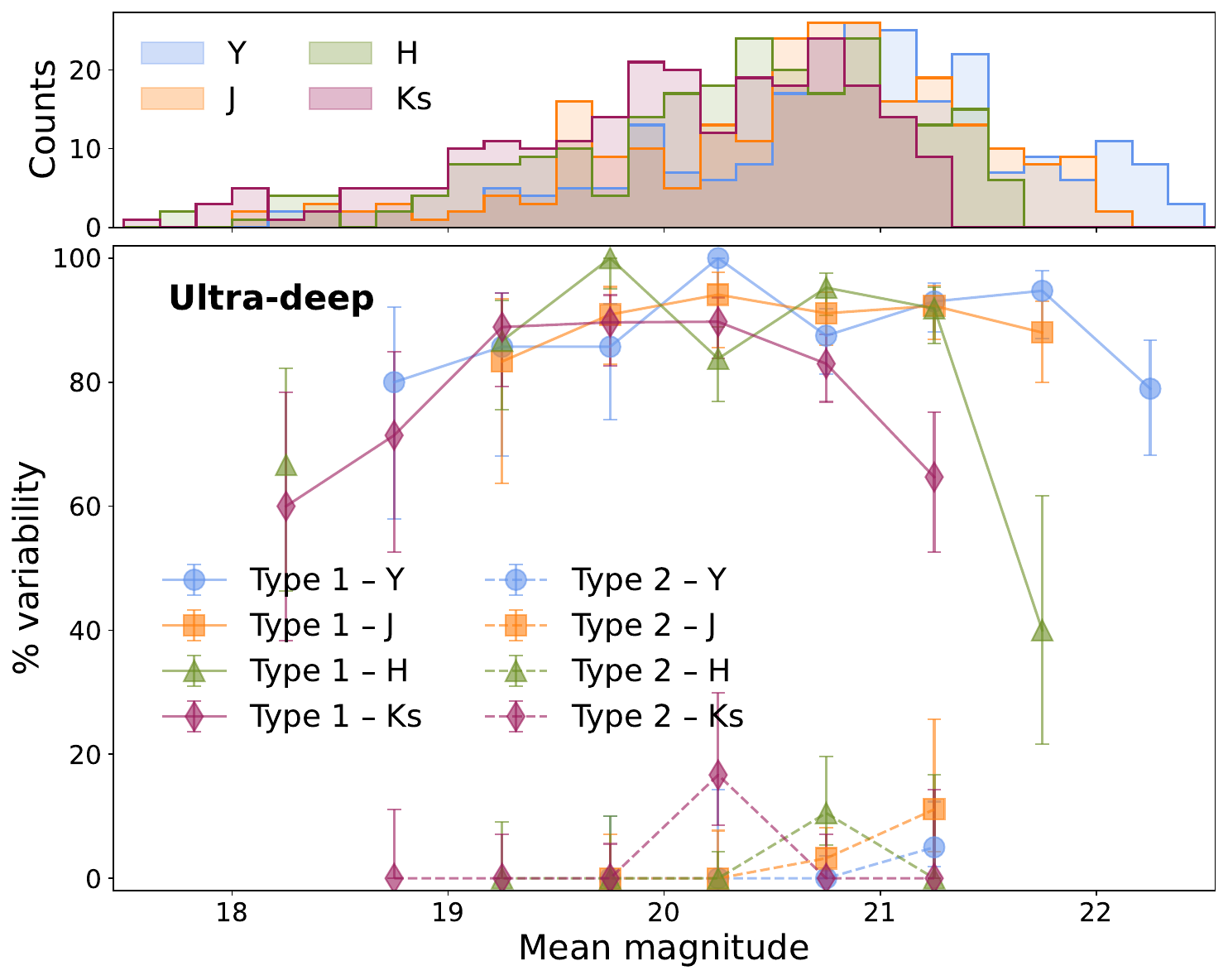}
  \end{subfigure}
  \begin{subfigure}
    \centering
    \includegraphics[width=\linewidth]{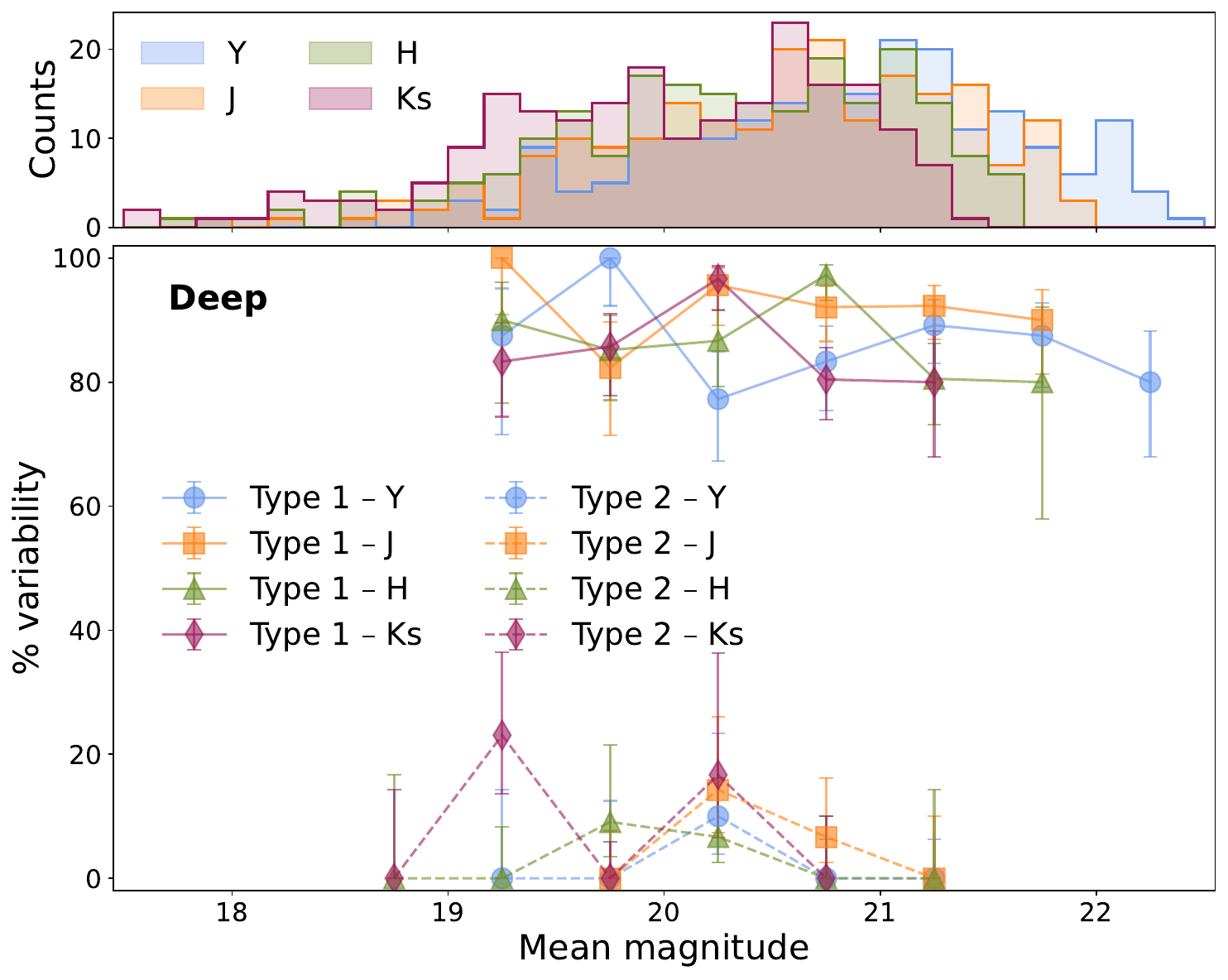}
  \end{subfigure}
\caption{
Fraction of variable sources as a function of observed magnitude in each NIR band ($YJHK_s$). The top panel corresponds to the ultra-deep stripes and the bottom panel to the deep stripes. In both cases, the upper subpanels show the magnitude distribution in the four bands. The lower subpanels display the variability fraction in each band, where different markers indicate the NIR bands (circles: $Y$, squares: $J$, triangles: $H$, diamonds: $K_s$), and solid and dashed lines distinguish type 1 and type 2 AGNs, respectively. Error bars represent the $1\sigma$ Wilson confidence intervals for each bin.
}
\label{fig:pct_var_mag}
\end{figure}

In this section we explore the potential dependence of the detectability of the variability on the apparent magnitude of the AGNs. For this purpose, we split each sample in magnitude bins, in each band, and calculate their  fraction of variable objects. The results are shown in Fig. \ref{fig:pct_var_mag} and the same data is tabulated in Tables \ref{table:var_mag_udeep} and \ref{table:var_mag_deep}.

\begin{table*}
\renewcommand{\arraystretch}{1.4}
\centering
\caption{
Variability fraction in bins of magnitude for type 1 and type 2 AGNs in the ultra-deep stripes for the four NIR bands ($YJHK_s$).}
\label{table:var_mag_udeep}
\resizebox{\textwidth}{!}{
\begin{tabular}{clccccccccc}
\hline

Magnitude bins (mag) &  & 18.0--18.5 & 18.5--19.0 & 19.0--19.5 & 19.5--20.0 & 20.0--20.5 & 20.5--21.0 & 21.0--21.5 & 21.5--22.0 & 22.0--22.5 \\
\hline\hline
\\[-1.0ex]
Type 1 
& $Y$  & -- & $80^{+12}_{-22}$ & $86^{+9}_{-18}$ & $86^{+7}_{-12}$ & $100^{+0}_{-6}$ & $87.5^{+4.3}_{-6.2}$ & $93.0^{+3.0}_{-4.9}$ & $94.7^{+3.2}_{-7.7}$ & $79^{+8}_{-11}$ \\
& $J$  & -- & -- & $83^{+10}_{-20}$ & $90.9^{+4.5}_{-8.0}$ & $94.1^{+3.6}_{-8.5}$ & $91.1^{+3.4}_{-5.2}$ & $92.3^{+3.3}_{-5.4}$ & $88.0^{+5.1}_{-8.0}$ & -- \\
& $H$  & $67^{+16}_{-20}$ & -- & $87^{+7}_{-11}$ & $100^{+0}_{-5}$ & $83.8^{+5.2}_{-6.9}$ & $95.2^{+2.4}_{-4.5}$ & $91.9^{+3.5}_{-5.7}$ & $40^{+22}_{-18}$ & -- \\
& $K_s$& $60^{+18}_{-22}$ & $71^{+14}_{-19}$ & $88.9^{+5.5}_{-9.5}$ & $89.7^{+4.4}_{-7.0}$ & $89.7^{+3.9}_{-5.9}$ & $83.0^{+4.8}_{-6.2}$ & $65^{+11}_{-12}$ & -- & -- \\ [1.0ex]

\hline
\\[-1.0ex]
Type 2 
& $Y$  & -- & -- & -- & $0^{+10}_{-0}$ & $0^{+14}_{-0}$ & $0^{+4}_{-0}$ & $5.0^{+7.4}_{-3.1}$ & -- & -- \\
& $J$  & -- & -- & -- & $0^{+7}_{-0}$ & $0^{+8}_{-0}$ & $3.2^{+4.9}_{-2.0}$ & $11^{+15}_{-7}$ & -- & -- \\
& $H$  & -- & -- & $0^{+9}_{-0}$ & $0^{+10}_{-0}$ & $0^{+4}_{-0}$ & $10.5^{+9.1}_{-5.2}$ & $0^{+17}_{-0}$ & -- & -- \\
& $K_s$& -- & $0^{+11}_{-0}$ & $0^{+7}_{-0}$ & $0^{+6}_{-0}$ & $17^{+13}_{-8}$ & $0^{+7}_{-0}$ & $0^{+14}_{-0}$ & -- & -- \\ [1.0ex]

\hline\hline
\end{tabular}
}
\tablefoot{Values are reported as percentages. The uncertainties correspond to the $1\sigma$ Wilson confidence intervals.}
\end{table*}

\begin{table*}
\renewcommand{\arraystretch}{1.4}
\centering
\caption{
Same as Table~\ref{table:var_mag_udeep}, but for the deep stripes.}
\label{table:var_mag_deep}
\resizebox{\textwidth}{!}{
\begin{tabular}{clccccccccc}
\hline
Magnitude bins (mag) &  & 18.0--18.5 & 18.5--19.0 & 19.0--19.5 & 19.5--20.0 & 20.0--20.5 & 20.5--21.0 & 21.0--21.5 & 21.5--22.0 & 22.0--22.5 \\
\hline\hline
\\[-1.0ex]
Type 1 
& $Y$  & -- & -- & $88^{+8}_{-16}$ & $100^{+0}_{-8}$ & $77^{+8}_{-10}$ & $83.3^{+5.7}_{-7.9}$ & $89.2^{+4.1}_{-6.2}$ & $87.5^{+5.3}_{-8.3}$ & $80^{+8}_{-12}$ \\
& $J$  & -- & -- & $100^{+0}_{-9}$ & $82^{+7}_{-11}$ & $95.7^{+2.7}_{-6.5}$ & $92.1^{+3.4}_{-5.5}$ & $92.3^{+3.3}_{-5.4}$ & $90.0^{+4.9}_{-8.7}$ & -- \\
& $H$  & -- & -- & $90^{+6}_{-13}$ & $85.2^{+5.6}_{-8.1}$ & $86.7^{+5.0}_{-7.4}$ & $97.3^{+1.7}_{-4.2}$ & $80.6^{+5.7}_{-7.4}$ & $80^{+12}_{-22}$ & -- \\
& $K_s$& -- & -- & $83.3^{+6.2}_{-8.9}$ & $85.7^{+5.4}_{-7.9}$ & $96.7^{+2.1}_{-5.1}$ & $80.4^{+5.2}_{-6.5}$ & $80^{+8}_{-12}$ & -- & -- \\ [1.0ex]

\hline
\\[-1.0ex]
Type 2 
& $Y$  & -- & -- & $0^{+14}_{-0}$ & $0^{+13}_{-0}$ & $10^{+13}_{-6}$ & $0^{+6}_{-0}$ & $0^{+6}_{-0}$ & -- & -- \\
& $J$  & -- & -- & -- & $0^{+8}_{-0}$ & $14^{+12}_{-7}$ & $6.7^{+9.5}_{-4.1}$ & $0^{+10}_{-0}$ & -- & -- \\
& $H$  & -- & $0^{+17}_{-0}$ & $0^{+8}_{-0}$ & $9^{+12}_{-6}$ & $6.7^{+9.5}_{-4.1}$ & $0^{+10}_{-0}$ & $0^{+14}_{-0}$ & -- & -- \\
& $K_s$& -- & $0^{+14}_{-0}$ & $23^{+13}_{-10}$ & $0^{+6}_{-0}$ & $17^{+20}_{-10}$ & $0^{+10}_{-0}$ & -- & -- & -- \\ [1.0ex]

\hline\hline
\end{tabular}
}
\end{table*}

\section{Examples of NIR and MIR light curves of variable type 2 AGNs}\label{appendix:lc}

Representative examples of NIR (UltraVISTA) and MIR (WISE) light curves for the three different classes of variable type 2 AGNs discussed in this work: disk-dominated, torus-dominated and highly obscured. Each panel displays the calibrated $YJHK_s$ light curves from UltraVISTA DR6, together with the W1 and W2 light curves from the unTimely Catalog Explorer \citep{Meisner2023}. 

\begin{figure}
  \centering
  \includegraphics[width=0.67\linewidth]{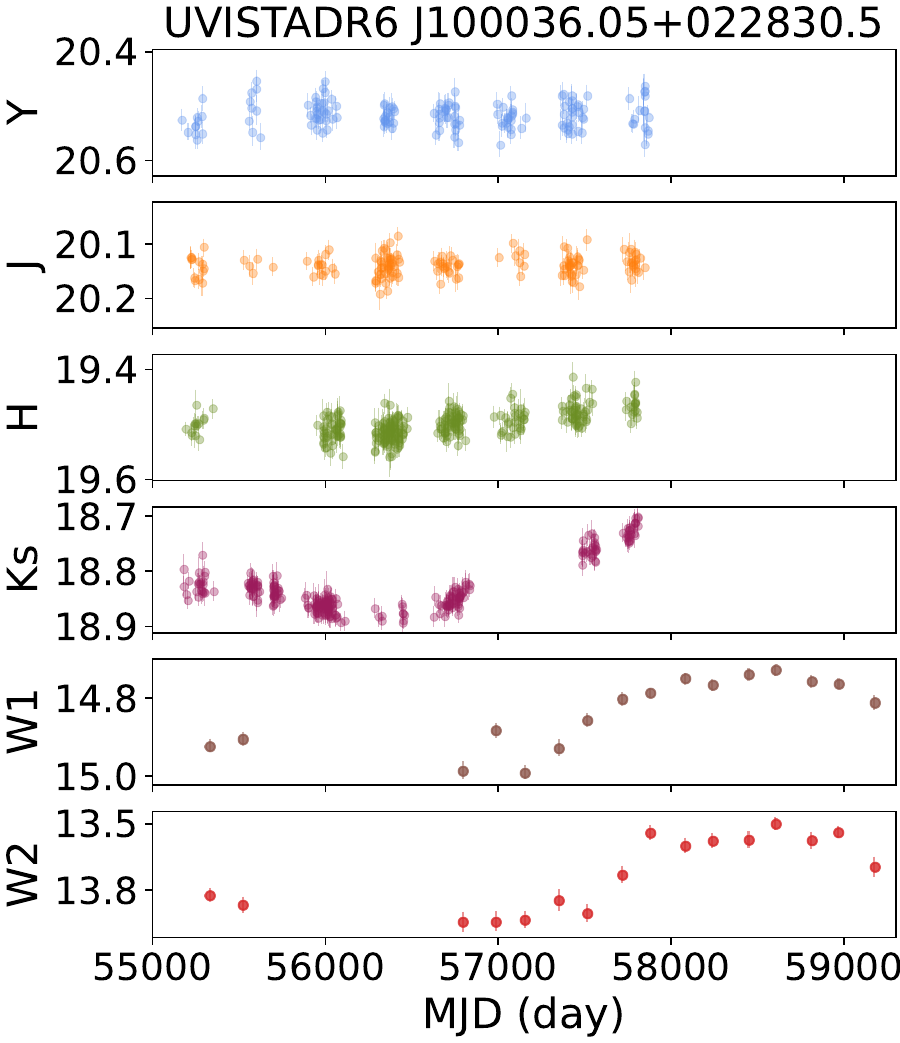}
  \caption{ 
NIR ($YJHK_s$) and MIR (W1, W2) light curves of the highly obscured type 2 XR II AGN at $z = 0.69$.}
  \label{fig:353742}
\end{figure}

\begin{figure}
  \centering
  \includegraphics[width=0.67\linewidth]{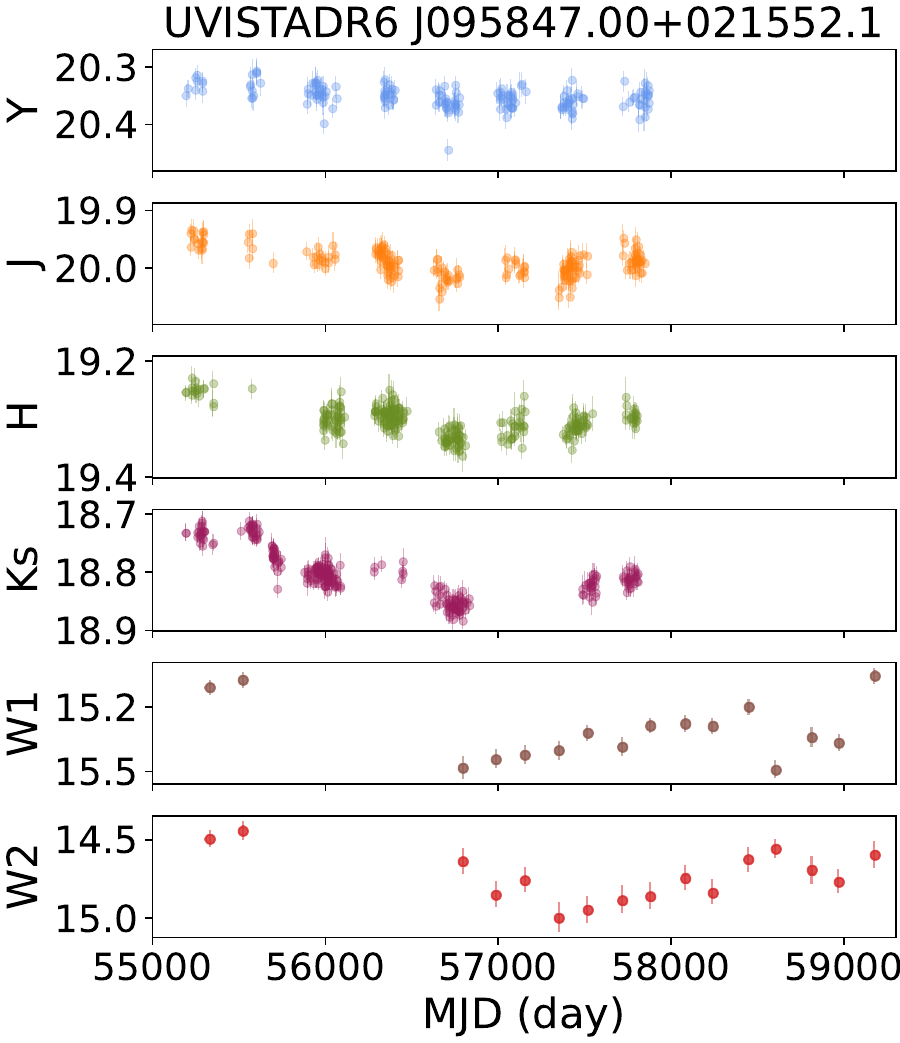}
  \caption{
NIR ($YJHK_s$) and MIR (W1, W2) light curves of the torus-dominated type 2 XR II AGN at $z = 0.55$.}
  \label{fig:262128}
\end{figure}

\begin{figure}
  \centering
  \includegraphics[width=0.67\linewidth]{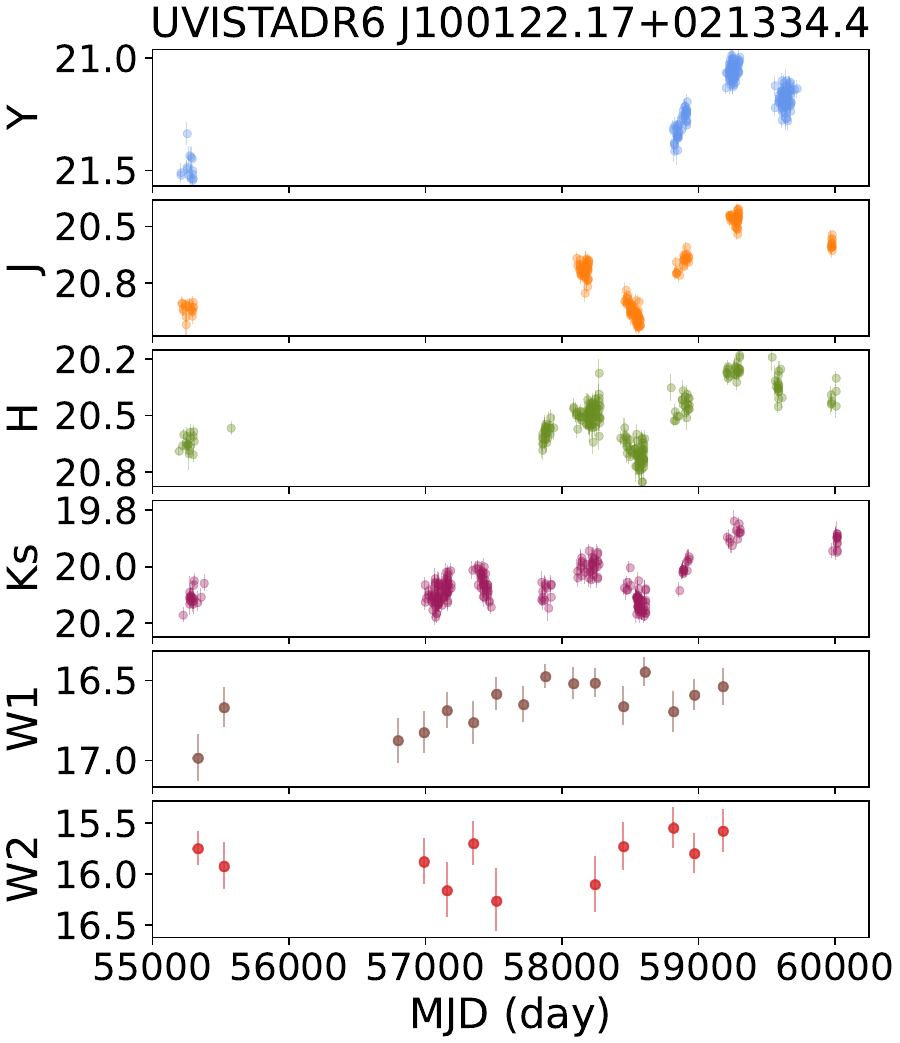}
  \caption{
NIR ($YJHK_s$) and MIR (W1, W2) light curves of the disk-dominated type 2 XR I AGN at $z = 0.89$.}
  \label{fig:114913}
\end{figure}

\section{X-ray and non--X-ray-detected type 2 AGNs}\label{appendix:xray_nonxray}

In this section we compare the redshift and magnitude distributions of X-ray-detected and non-X-ray type 2 AGNs. These comparisons are used to assess whether differences in their observed properties could affect the variability analysis.

\begin{figure*}
  \centering
  \includegraphics[width=\linewidth]{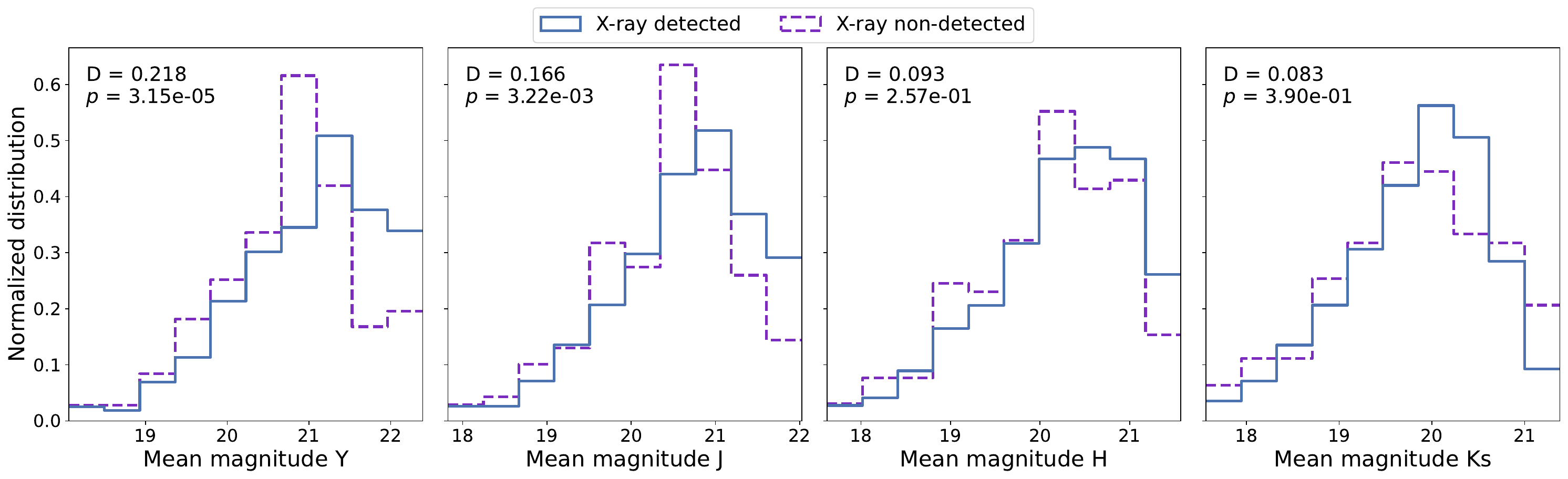}
  \caption{Magnitude distributions of the X-ray-detected (blue solid lines) and X-ray nondetected (purple dashed lines) type 2 AGNs in the four NIR bands. The K-S distance ($D$) and $p$-value are indicated in each panel.}
  \label{fig:xray_nonxray_mag}
\end{figure*}

\begin{figure}
  \centering
  \includegraphics[width=\linewidth]{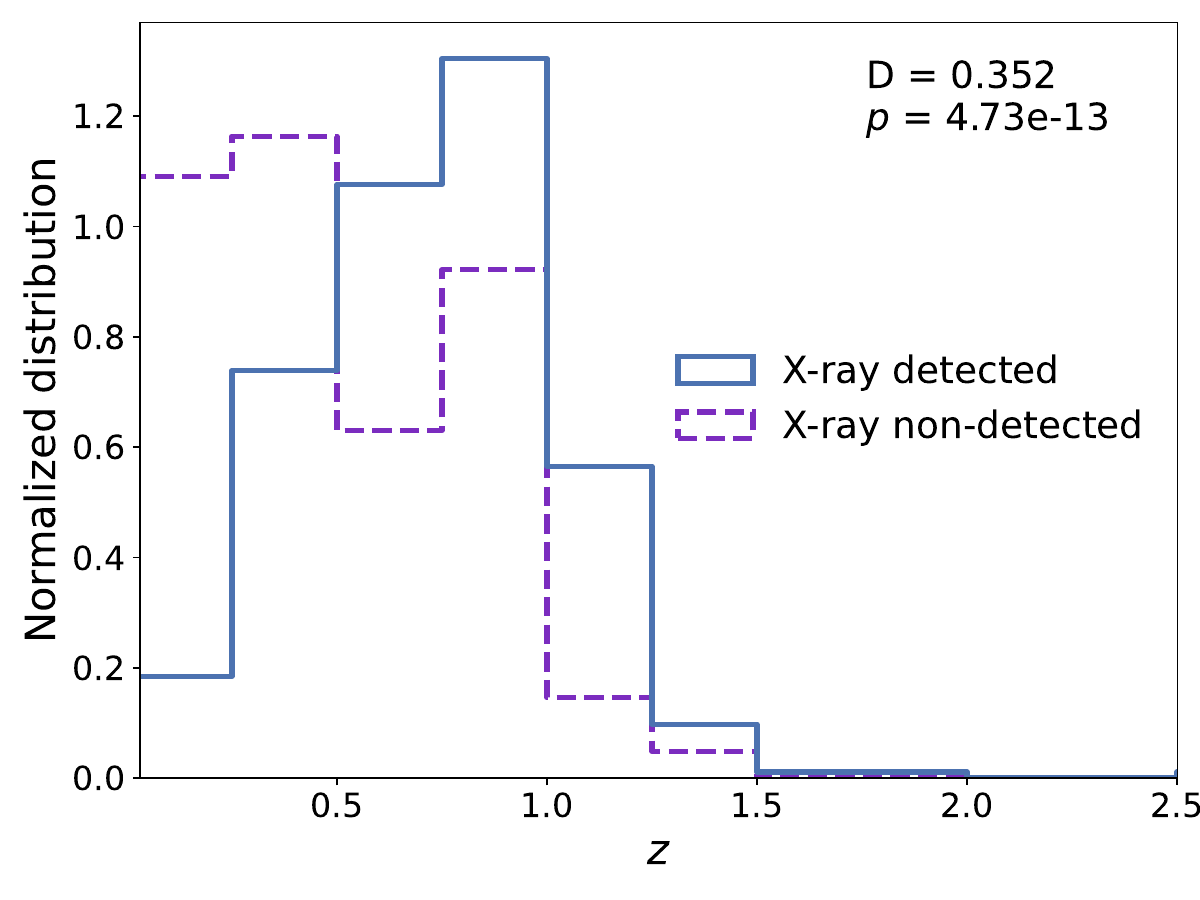}
  \caption{Redshift distributions of the X-ray-detected (blue solid lines) and X-ray nondetected (purple dashed lines) type 2 AGNs. The K-S distance ($D$) and $p$-value are indicated in the figure. The median redshifts of the X-ray-detected and X-ray nondetected samples are 0.75 and 0.46, respectively.}
  \label{fig:xray_nonxray_z}
\end{figure}

\end{document}